\documentclass[%
 reprint,
 amsmath,amssymb,
 aps,
]{revtex4-2}

\usepackage{graphicx}
\usepackage{dcolumn}
\usepackage{bm}
\usepackage{comment}
\usepackage{xcolor}
\usepackage[normalem]{ulem}
\usepackage[caption=false]{subfig}

\usepackage{multirow}
\usepackage{array}

\usepackage{tabularx}
\usepackage{graphicx}
\usepackage{amsmath}
\usepackage{longtable}
\usepackage{url}
\usepackage{verbatim}
\usepackage{lipsum}
\usepackage{mathtools, nccmath}

\begin{document}

\preprint{APS/123-QED}

\title{Universal trimers with $p$-wave interactions and the faux-Efimov effect}

\author{Yu-Hsin Chen}
\email{yuhsinchen@mail.ntou.edu.tw}
\affiliation{Department of Marine Engineering, National Taiwan Ocean University, Keelung 20224, Taiwan}
\author{Chris H. Greene}
\email{chgreene@purdue.edu}
\affiliation{Department of Physics and Astronomy, Purdue University, West Lafayette, Indiana 47907 USA}
\affiliation{Purdue Quantum Science and Engineering Institute, Purdue University, West Lafayette, Indiana 47907 USA}

\date{\today}

\begin{abstract}
An unusual class of $p$-wave universal trimers with symmetry $L^{\Pi}=1^{\pm}$ is identified, for both a two-component fermionic trimer with $s$- and $p$-wave scattering length close to unitarity and for a one-component fermionic trimer at $p$-wave unitarity.  Moreover, fermionic trimers made of atoms with two internal spin components are found for $L^{\Pi}=1^{\pm}$, when the $p$-wave interaction between spin-up and spin-down fermions is close to unitarity and/or when the interaction between two spin-up fermions is close to the $p$-wave unitary limit. The universality of these $p$-wave universal trimers is tested here by considering van der Waals interactions in a Lennard-Jones potential with different numbers of two-body bound states; our calculations also determine the value of the scattering volume or length where the trimer state hits zero energy and can be observed as a recombination resonance. The faux-Efimov effect appears with trimer symmetry $L^{\Pi}=1^{-}$ when the two fermion interactions are close to $p$-wave unitarity and the lowest $1/R^2$ coefficient gets modified, thereby altering the usual Wigner threshold law for inelastic processes involving 3-body continuum channels.
\end{abstract}

\maketitle


\section{Introduction}

In recent decades, ultracold quantum gases have made major strides by implementing tunability of the two-body interaction, as in the use of Fano-Feshbach resonances to manipulate the $s$-wave scattering length or the $p$-wave scattering volume in quantum gases\cite{chin2010feshbach, Greene2017RMP}. Through the use of these techniques, many fascinating phenomena exist in many-body environments controlled by ultracold collisions. For example, Bose-Einstein condensation was observed with lithium\cite{bradley1997bose,zwierlein2003observation,schreck2001quasipure} or potassium atoms\cite{greiner2003emergence} and the BCS-BEC crossover has been extensively researched. When the $s$-wave scattering length is positive and large ($a_s > 0$), universal $s$-wave molecules exist, and when the $s$-wave scattering length is negative and large ($a_s < 0$) between two different spin fermions, there are long range Cooper pairs in the gas\cite{blume2007universal,von2007spectrum,strecker2003conversion,chen2005bcs}. 

For identical bosons with total angular momentum $J^{\Pi}=0^+$, when the $s$-wave scattering length goes to negative infinity ($a_s \rightarrow -\infty$) for each pair of the bosons, an infinity of geometrical energy levels of trimer bound states exists. The three-body adiabatic potential curve is attractive at long-range, which is the now well-known Efimov effect\cite{efimov1970energy,efimov1973energy,naidonEfimovPhysicsReview2017,braaten2006universality, WangAAMOPHYS2013,dincaoFewbodyPhysicsResonantly2018}.

For trimers consisting of single-component fermions, a $p$-wave Efimov effect in the symmetry $L^{\Pi}=1^{-}$ was initially predicted by Macek and Sternberg \textit{et al.}\cite{macek2006properties}. However, that prediction was disproven by effective field theory; that prediction of a $p$-wave Efimov effect in zero-range theory turned out to be unphysical since it led to energy eigenstates having negative probability\cite{braaten2006universality,nishida2012impossibility}. In the two-component fermionic equal mass trimers, there are no known symmetries having an Efimov effect at $s$-wave unitarity. For the unequal mass system of two spin-up fermions and one spin-down fermion, the Efimov effect occurs when the mass ratio of spin-up fermions to a spin-down fermion is greater than $\approx 13.6$\cite{braaten2012renormalization}. Braaten \textit{et al.} also found an Efimov effect at $p$-wave unitarity when there is a strong $p$-wave interaction between the third equal mass particle and each of the two identical fermions, for symmetries $L^{\Pi}=0^+, \, 1^{\pm}$ and $2^+$ or two identical bosons with $L^{\Pi}=1^{+}$\cite{braaten2012renormalization}. Nevertheless, these equal mass cases of a predicted fermionic Efimov effect have all been proven by Nishida \textit{et al.}\cite{nishida2012impossibility} to be unphysical. The $s$-wave scattering length goes to a large positive between spin-up and spin-down Fermi gases with angular momentum $L^{\Pi}=1^-$.

In the system of two spin-up fermions with mass $m_2$ and one spin-down fermion (or boson) with mass $m_1$, there are two $s$-wave universal trimer states when the mass ratio $ 12.3 \lesssim m_2/m_1 \lesssim 13.6$.  Moreover, there is one $s$-wave universal trimer state with mass ratio $ 8.2 \lesssim m_2/m_1 \lesssim 12.3$ and there is no $s$-wave universal trimer with mass ratio $m_2/m_1 	\lesssim 8.2$ \cite{kartavtsev2007low}. For these universal trimer cases, we abbreviate them in the following as the ``KM trimer(s)." 
Other cases of unequal mass universal trimers have been predicted, as in \cite{kartavtsev2007low}.
Moreover, in a recent study, Naidon {\it et al.} found a universal trimer state that occurs when the heavy/light mass ratio is smaller than 8.2, all the way down to the equal mass regime, when the $p$-wave interaction between two spin-up fermion\cite{naidon2022shallow} is tuned near unitarity. 

Our explorations consider additional scenarios involving three equal mass fermions, with either a single-spin component or with two components,  and we find a trimer state for symmetry $L^{\Pi}=1^\pm$. For instance, when the two spin-up fermion interaction is tuned close to the $p$-wave unitary limit, while the interaction between the spin-up and spin-down fermions is fixed at the $s$-wave unitary limit, a universal trimer state was found for three equal mass fermions with symmetry $L^{\Pi}=1^-$\cite{chen2022p}. Moreover, there is initial evidence for an  Efimov effect that appears in the {\it Born-Oppenheimer} potential curves but which is {\it not present} in the more relevant and appropriate {\it adiabatic} potential curves that include the diagonal adiabatic correction (also denoted the Born-Huang correction in the literature\cite{born1996dynamical}). This phenomenon has been denoted previously and throughout the present article as the {\it faux-Efimov effect}\cite{chen2022efimov}; i.e., there would be an Efimov effect present in the Born-Oppenheimer potential curves but ultimately no Efimov effect showing in the effective adiabatic potential\cite{chen2022efimov} with trimer angular momentum $L^{\Pi}=1^-$. Other types of $p$-wave universal trimers have been identified and are introduced in this article, which arise when fermions are in either the same or different spin states. Examples also arise either with or without a doubly-degenerate faux-Efimov effect for the system of three equal mass fermions. 

Furthermore, our broadened understanding of Efimov physics also demonstrates that the coefficient of $R^{-2}$ in the lowest continuum adiabatic potential curve at large hyperradii, which determines a class of threshold exponents, becomes modified when each pair of boson or fermion or mixed quantum gases approach either the $s$-wave\cite{werner2006unitary,higgins2020nonresonant} or the higher partial wave ($p$-wave) unitary limit\cite{chen2022efimov,d2005scattering,higgins2022three}. The lowest three-body continuum adiabatic potential curve is significant for the three-body recombination rate, as the coefficient of $1/R^2$ controls the exponent in the three-body continuum Wigner threshold law for all inelastic processes\cite{suno2003recombination,d2005scattering,d2008ultracold,d2008suppression,chen2022p}. For example, for the two-component fermionic trimer at large and negative $s$-wave scattering length ($a_s \ll 0$), the three-body recombination rate is proportional to $\lvert a_s \rvert^{2.455}$\cite{d2005scattering}. For large and positive $s$-wave scattering length ($a_s \gg 0$) on the other hand, the recombination rate is proportional to $a_s^6$\cite{petrov2003three,d2005scattering,ji2022stability}. In previous studies, the interaction between spin-up and spin-down fermions was fixed at the $s$-wave unitary limit and with the $p$-wave interaction between two spin-up fermions tuned through large values of the $p$-wave scattering volume ($V_p$), the three-body recombination rate ($K_3$) was predicted to scale as $K_3 \propto \lvert V_p \rvert^{1.182}$\cite{chen2022p}.  In the ultracold many-body problem, therefore, with each pair of particles interacting near either the $s$-wave or $p$-wave unitary limit, Efimov physics plays a key role to control the recombination and/or dissociation rate for Wigner threshold law, which can determine the experimental lifetime for such an ultracold quantum gas of fermions. 

In section \ref{sec:s-wave}, a different type of $p$-wave universal trimer will be introduced, which is the two-component fermion at simultaneous $s$- and $p$-wave unitary limits. Both the faux-Efimov effect and a $p$-wave universal trimer have been found to emerge with trimer symmetry $L^{\Pi}=1^-$. The adiabatic potential curves from first to third $p$-wave pole are presented to enable more detailed studies, and the universality of the trimer state has been tested by using the derivative of the same of all eigenphase shifts with various $p$-wave poles. Also, the key coefficient of $R^{-2}$ in the asymptotic 3-body adiabatic potential curves for different angular momentum $L^{\Pi}$ is discussed, with comparisons depending on whether the interaction between each pair of fermions is near the $s$- and/or $p$-wave unitary limit.

In section \ref{sec:p-wave}, cases with trimers having only their pairwise $p$-wave interactions near unitarity are discussed. The Efimov-physics-modified coefficient of $R^{-2}$ in the asymptotic adiabatic potential curve at $p$-wave unitary limit has been determined and $p$-wave universal trimers with different trimer symmetries are discussed. In subsection \ref{subsec:p-wave_only_opposite}, the $p$-wave interaction is only large between the spin-up and spin-down fermions, while the two spin-up fermions have weak interaction. In subsection \ref{subsec:p-wave_all}, the interaction between each pair of fermion has been tuned to $p$-wave unitarity. A remarkable case of doubly-degenerate faux-Efimov and $p$-wave universal channels\cite{chen2022efimov} is introduced. 

In section \ref{sec:three p-wave}, the $p$-wave universal trimer for three equal mass and spin-up fermions at the $p$-wave unitary limit with $L^{\Pi}=1^{\pm}$ will be discussed. The lowest 60 adiabatic potential curves are shown and the corresponding scattering eigenphase shifts are calculated in order to identify low-lying resonance energies and decay widths. 

\section{Method}
The adiabatic hyperspherical representation has a strong track record in describing few-body interactions and collisional phenomena and is used here to analyze the three-body quantum problem\cite{suno2003recombination, Rittenhouse-2011JPB, Greene2017RMP}. The three-body adiabatic equation at fixed hyperradius $R$ for three equal mass fermions (or any other equal mass particles) can be written as 

\begin{equation}\label{eq:adSE}
    H_{\text{ad}}(R;\Omega)\Phi_\nu(R;\Omega)=U_\nu(R)\Phi_\nu(R;\Omega),
\end{equation}
whose eigensolutions are $R$-dependent and obtained by diagonalizing the adiabatic Hamiltonian matrix in a suitable basis set. Here we denote the eigenvalues $U_\nu (R)$ as the {\it Born-Oppenheimer potential curves}, and the eigenfunctions $\Phi_\nu(R;\Omega)$ are the corresponding channel functions. $\Omega$ represents the five hyperangles $\Omega \equiv (\theta, \varphi, \alpha, \beta, \gamma)$ plus any relevant spin degrees of freedom. The adiabatic Hamiltonian contains all hyperangular dependence and interactions, which can be defined as\cite{rittenhouse2011hyperspherical}
\begin{equation}\label{eq:Had}
    H_{\text{ad}}(R;\Omega) \equiv \frac{\hbar^2\Lambda^2(\theta,\varphi)}{2\mu_{3b}R^2}+\frac{15\hbar^2}{2\mu_{3b}R^2}+V(R,\theta,\varphi).
\end{equation}
where the $\mu_{3b}$ is the three-body reduced mass and $\Lambda^2(\theta,\varphi)$ is called the ``grand angular-momentum operator''\cite{kendrick1999hyperspherical}. The three-body interaction potential $V(R,\theta,\varphi)$ is taken to be a sum of two-body interactions $V(R,\theta,\varphi)=v_{1}(r_{23})+v_{2}(r_{31})+v_{3}(r_{12})$. The interparticle distance $r_{ij} $ can be written in these coordinates as
\begin{equation}
    r_{ij}=2^{-1/2}d_{ij}R\left[1+\sin\theta\cos(\varphi+\varphi_{ij})\right]^{1/2},
\end{equation}
with $\varphi_{12}=2\arctan(m_2/\mu_{3b})$, $\varphi_{23}=0$ and $\varphi_{31}=-2\arctan(m_3/\mu_{3b})$. There are three two-body coalescence points for three equal mass systems at $\varphi=\pi/3$, $\pi$, and $5\pi/3$. The two-body potential is the following, e.g., in the case of the Lennard-Jones potential\cite{wang2012origin} 

\begin{equation}\label{eq:LJpotential}
    v_i(r)=-\frac{C_{6}}{r^{6}}\left(1-\frac{\lambda_n^{6}}{r^{6}}\right).
\end{equation}

In this article, in our chosen set of van der Waals units, the $C_6$ coefficient is set at $16 \, r_{\text{vdW}}^{6}E_{\text{vdW}}$ where $r_{\text{vdW}}$ is the van der Waals length $r_{\text{vdW}}\equiv(m C_6/\hbar^2)^{1/4}/2$, ($\mu=m/2$ is the two-body reduced mass here) and $E_{\text{vdW}}$ is the van der Waals energy unit, $E_{\text{vdW}}\equiv \hbar^2/(2\mu r_{\text{vdW}}^2)$. The parameter $\lambda_n$ can be adjusted to produce any desired $s$-wave scattering length or $p$-wave scattering volume for a chosen pair of fermions. Here the two-body $s$-wave scattering length and $p$-wave scattering volume can be extracted from low energy $s-$ and $p-$wave solutions of the relative radial Schr\"odinger equation using the form:
\begin{equation}\label{eq:asVpdef}
    k^{2L+1} \cot(\delta_L) = -1/{a_L}^{2L+1}+\frac{1}{2} r_L k^2.
\end{equation}
Here $a_0(\equiv a_s)$ is the $s$-wave scattering length, and $a_1(\equiv a_p)$ is the $p$-wave scattering length, $\delta_{0}(k)$ is the $s$-wave scattering phase shift, and $\delta_{1}(k)$ is the $p$-wave scattering phase shift; the corresponding effective ranges are $r_0$ and $r_1$, with $k$ the wave number. The $s$-wave scattering length and $p$-wave scattering volume can be represented as, 
\begin{align}
    &a_s=-\lim_{k \to 0}\left[\frac{\tan\delta_0(k)}{k}\right], \label{eq:asVpdef2}\\
    &V_p=-\lim_{k \to 0}\left[\frac{\tan\delta_1(k)}{k^3}\right]\equiv a_p^3.\label{eq:asVpdef3}
 \end{align}
 Note that Eqs.\ref{eq:asVpdef}-\ref{eq:asVpdef3} are still applicable for a long range van der Waals tail, provided $L \leq 1$.

The Born-Oppenheimer potential and nonadiabatic couplings are obtained by solving Eq.(\ref{eq:adSE}) for fixed values of the hyperradius $R$. For each $R$, the set of channel functions $\Phi_\nu(R;\Omega)$ are orthogonal,
\begin{equation}\label{eq:Phi-orth}
    \int d\Omega \Phi_\mu(R;\Omega)^* \Phi_\nu(R;\Omega)=\delta_{\mu\nu},
\end{equation}
and complete
\begin{equation}
    \sum_\tau \Phi_\tau(R;\Omega) \Phi_\tau(R;\Omega')^* = \delta(\Omega-\Omega').
\end{equation}

The adiabatic representation then expands the full desired wave function $\psi_E(R;\Omega)$ in a truncated subset of the complete orthonormal set of hyperangular eigenfunctions $\Phi_\nu(R;\Omega)$, each multiplied by a corresponding radial wave function $F_{\nu E}(R)$ to be determined\cite{suno2009three}:
\begin{equation}\label{eq:psiE}
    \Psi_E(R;\Omega)=R^{-5/2}\sum_{\nu = 0}^{\infty}F_\nu^E(R)\Phi_\nu (R;\Omega).
\end{equation}
In order to diagonalize the adiabatic Hamiltonian Eq.(\ref{eq:adSE}), our treatment follows the standard route that expands the channel function into Wigner $D$ functions, which rotate from the laboratory frame into a body-fixed frame coordinate system:
\begin{equation}\label{eq:Phiexpand}
    \Phi_\nu(R;\Omega)=\sum_K^L \phi_{K\nu}(R;\theta,\varphi)D_{KM}^L(\alpha, \beta, \gamma).
\end{equation}
The quantum numbers $K$ and $M$ represent the projection of the orbital angular momentum operator $\mathbf{L}$
onto the body-fixed and space-fixed $z$-axes, respectively. The parity can be written as $\Pi=(-1)^K$, which gives $\hat{\Pi}D_{KM}^L=(-1)^K D_{KM}^L$\cite{kendrick1999hyperspherical}. As is known from the nuclear and atomic collisions literature, a given symmetry is called ``parity favored'' if $\Pi=(-1)^L$, and is called ``parity unfavored'' otherwise\cite{fano1964exclusion}. In the present system, for the parity favored case, $L-K$ is even and $K$ takes the values $L,L-2,...,-(L-2),-L$. For the parity unfavored case, $L-K$ is odd, and $K$ takes the values $L-1,L-3,...,-(L-3),-(L-1)$. 

The three-body Schr\"{o}dinger equation can be written using modified Smith-Whitten hyperspherical coordinates\cite{suno2002three, whitten1968symmetric, wang2012origin} and after the rescaling transformation $\psi_E\equiv R^{5/2}\Psi$, the equation becomes\cite{rittenhouse2011hyperspherical}:
\begin{align}\label{eq:totalSE2}
    &\left[-\frac{\hbar^2}{2\mu_{3b}}\frac{\partial^2}{\partial R^2}+\frac{15\hbar^2}{8\mu_{3b}R^2}+\frac{\hbar^2\Lambda^2(\theta,\varphi)}{2\mu_{3b}R^2}+V(R,\theta,\varphi)\right]\psi_E \nonumber\\ 
    &=E\psi_E.
\end{align}
Substitution of the truncated expansion Eq.(\ref{eq:psiE})
into the Schr\"{o}dinger equation Eq.(\ref{eq:totalSE2}) leads to a set of one-dimensional coupled hyperradial differential equations:
\begin{align}\label{eq:1D-SE}
    &\left[-\frac{\hbar^2}{2\mu_{3b}}\frac{d^2}{dR^2}+W_\nu(R)-E \right]F_{\nu}^E(R) \nonumber\\
    &+\sum_{\nu'\neq\nu} \left(-\frac{\hbar^2}{2\mu_{3b}} \right)  \left[2P_{\nu\nu'}(R)\frac{d}{dR}+Q_{\nu\nu'}(R)\right]F_{\nu'}^E(R)=0
\end{align}

In the above expression, $E$ is the total energy and $W_\nu(R)$ is the effective {\it adiabatic potential} in channel $\nu$:
\begin{equation}
    W_\nu(R) \equiv U_\nu(R)-\frac{\hbar^2}{2\mu_{3b}}Q_{\nu\nu}(R).
\end{equation}
The adiabatic potential includes the so-called `diagonal correction' or `Born-Huang correction' $Q_{\nu \nu}(R)$, while we sometimes refer instead to the uncorrected `Born-Oppenheimer' potential curve $U_\nu(R)$.
The nonadiabatic coupling matrices $P_{\nu\nu'}(R)$ and $Q_{\nu\nu'}(R)$ are defined as \begin{align}
	&P_{\nu\nu'}(R) \equiv \int d\Omega \Phi_{\nu}^*(R;\Omega)\frac{\partial}{\partial R}\Phi_{\nu'}(R;\Omega), \\
	&Q_{\nu\nu'}(R) \equiv \int d\Omega \Phi_{\nu}^*(R;\Omega)\frac{\partial^{2}}{\partial R^{2}}\Phi_{\nu'}(R;\Omega).
\end{align} 
By taking the hyperradial derivative of Eq.(\ref{eq:Phi-orth}), one readily sees that the $P_{\nu\nu'}(R)$ matrix is anti-symmetric, since the usual phase convention is adopted where these matrix elements are all real. If one next differentiates the above definition of $P_{\nu\nu'}(R)$, this gives a relation between $P_{\nu\nu'}(R)$ and $Q_{\nu\nu'}(R)$, namely 
\begin{equation}
    \frac{d}{dR}P_{\nu\nu'}(R)=-P_{\nu\nu'}^2(R)+Q_{\nu\nu'}(R),
\end{equation}
where the $P_{\nu\nu'}^2(R)$ matrix can be expressed as 
\begin{equation}
    P_{\nu\nu'}^2(R)=\int d\Omega \frac{\partial}{\partial R}\Phi_{\nu}^*(R;\Omega)\frac{\partial}{\partial R}\Phi_{\nu'}(R;\Omega).
\end{equation}
Some other properties of the $P_{\nu\nu'}(R)$ matrix are: $P_{\nu\nu'}=-P_{\nu'\nu}$ and $P_{\nu\nu'}^2=P_{\nu'\nu}^2$, which leads to $P_{\nu\nu}=0$ and $Q_{\nu\nu}=-P_{\nu\nu}^2$. 
According to the above properties of $P_{\nu\nu'}(R)$, only the $P_{\nu\nu'}^2(R)$ component of $Q_{\nu\nu'}$ is needed in order to solve the one-dimensional radial coupled equation Eq.(\ref{eq:1D-SE}).

In the absence of Coulomb interactions, the asymptotic effective adiabatic potentials in the 3-body continuum are accurately characterized at $R\rightarrow\infty$ as\cite{nielsen2001three}
\begin{equation}\label{eq:wasym}
	W_{\nu}(R) \rightarrow \frac{\hbar^2 l_{e,\nu}(l_{e,\nu}+1)}{2\mu_{3b} R^{2}}=\frac{\hbar^2[\lambda_\nu(\lambda_\nu+4)+15/4]}{2\mu_{3b}R^2}.
\end{equation}
where $l_{e,\nu}$ controls the effective angular momentum barrier of the three free asymptotic particles at large hyperradius ($R\rightarrow\infty$) and $\nu$ represents the $\nu$-th channel. In some references, the $l_{e,\nu}$ value has been recast as $l_{e,\nu}=\lambda_\nu+3/2$. The $l_{e,\nu}$ value also determines the
scaling law of the three-body recombination rate and the squared scattering matrix element, through the Wigner threshold law $\lvert S_{j\leftarrow i}^{L\Pi} \lvert^2 \propto {k_i}^{2l_{e,i}+1}$\cite{esry2001threshold}. The long-range effective adiabatic potential curves representing any atom plus dimer channel can be represented asympotically as
\begin{equation}\label{eq:atomdimer}
    W_{\nu}(R)\rightarrow E_{\nu l}+\frac{\hbar^2 l_\nu'(l_\nu'+1)}{2\mu R^2}.
\end{equation}
Here $E_{\nu l}$ is the rovibrational dimer energy (two-body bound state), $l$ represents the dimer angular momentum, $\nu$ is the $\nu$-th channel, and $l_\nu'$ is the angular momentum of the third particle relative to the dimer.

\section{Two-component Fermi trimers at $S$-wave unitarity}\label{sec:s-wave}

Recently, a previously unknown type of $p$-wave universal trimer was predicted\cite{chen2022p} to exist for three equal mass fermionic atoms at the $s$-wave unitary limit, in particular for the special value of the total angular momentum and parity, $L^{\Pi}=1^{-}$. Previously an $s$-wave universal trimer with a different mass ratio between the spin-up and spin-down fermions was comprehensively researched by Kartavtsev {\it et al.} \cite{kartavtsev2007low,kartavtsev2014recent}.   That system shows an Efimov effect when the mass ratio of majority-spin (up) fermions ($m_{\uparrow}$) divided by the mass of the lone spin-down fermion ($m_{\downarrow}$) is larger than $13.606$ at $s$-wave unitary limit, namely, $a_{s}\rightarrow\infty$ with angular momentum $L^{\Pi}=1^{-}$. Moreover, a first $s$-wave universal trimer occurs when the mass ratio obeys $m_{\uparrow}/m_{\downarrow} > 8.172$, and a second one appears at mass ratio $m_{\uparrow}/m_{\downarrow} > 12.917$\cite{endo2012crossover}. 

Our studies demonstrate that, for the three equal mass fermion system (i.e. with the mass ratio $m_{\uparrow}/m_{\downarrow} =1$), a $p$-wave universal trimer will emerge when the $p$-wave attractive interaction between spin-up fermions is strengthened, while the different-spin fermions still retain their near-infinite value of the $s$-wave scattering length value, $a_s\rightarrow\infty$. Our prediction is that another type of fermion trimer with three equal mass particles should exist in the symmetry $L^{\Pi}=1^{-}$. Based on our previous study\cite{chen2022efimov} and other previous research by other groups, an $s$-wave unitary channel exists for a two-component fermionic trimer at the $s$-wave unitary limit with angular momentum $L^{\Pi}=1^{-}$, for which the $l_e$ value in the $l_e(l_e+1)$ coefficient of $1/(2\mu R^2)$ in the lowest continuum channel is reduced (by the Efimov physics) to the value $l_e=1.272$\cite{werner2006unitary,blume2007universal,higgins2020nonresonant}. If there is no interaction between each pair of fermions or the hyperradius is large and the potential converges to zero energy at infinity, the $l_e$ value of the lowest continuum channels is $5/2$. 

At the same time, the interaction between two spin-up fermionic atoms at the $p$-wave unitarity limit has a modified lowest continuum channel where the $l_e$ value approaches to $0$ asymptotically, which we have denoted as a {\it $p$-wave unitary channel}. Owing to the presence of this novel channel, the new type of trimer can be created when the interaction between each pair of different-spin fermions is set at the $s$-wave unitary limit with symmetry $L^{\Pi}=1^{-}$. 
\begin{figure}[htbp]
    \centering
    \includegraphics[width=8.5cm]{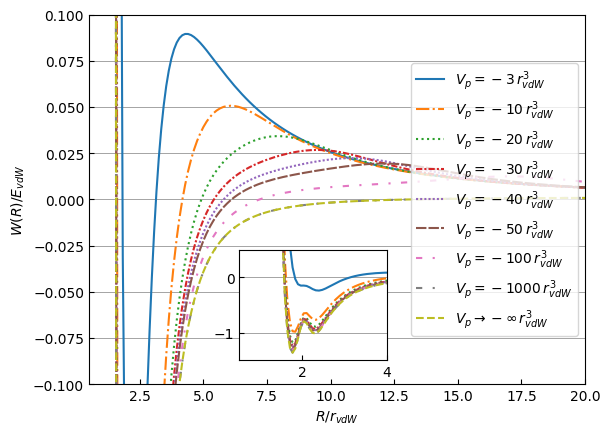}
    \caption{(color online). Shown are the lowest effective adiabatic potential curves versus hyperradius for several different $p$-wave scattering volumes ($V_{p}$) in the van der Waals length unit $r_{\text{vdW}}$, for the system $\downarrow\uparrow\uparrow$ with the interaction between spin-up and spin-down fermion set at the $s$-wave unitary limit with $L^{\Pi}=1^-$. As $\lvert V_p \rvert$ gets larger, the potential curve should approach the three-body threshold whose value closes the $l_e=0$ in the adiabatic potential energy curve. The inset plots the behavior of depth of effective adiabatic potential curves versus the various scattering volume $V_p$.}
    \label{fig:spsVp2}
\end{figure}

Fig.\ref{fig:spsVp2} shows the effect of tuning the $p$-wave scattering volume characterizing the interaction between the two spin-up fermions in a two-component fermionic trimer for the symmetry $L^{\Pi}=1^-$.  The $s$-wave interaction between the spin-up and spin-down fermions is fixed at $a_{s}\rightarrow \infty$. If the $V_p$ is negative and small, namely very weak interaction between the pair of spin-up fermion, no trimer states exist, consistent with the known inequality in the mass ratio that is necessary for formation of the KM trimer\cite{kartavtsev2007low,kartavtsev2014recent}. 

\onecolumngrid

\begin{figure}[htbp]
    \centering
    \subfloat[\label{fig:sps2ndpwave}]
    {\includegraphics[width=8.5cm]{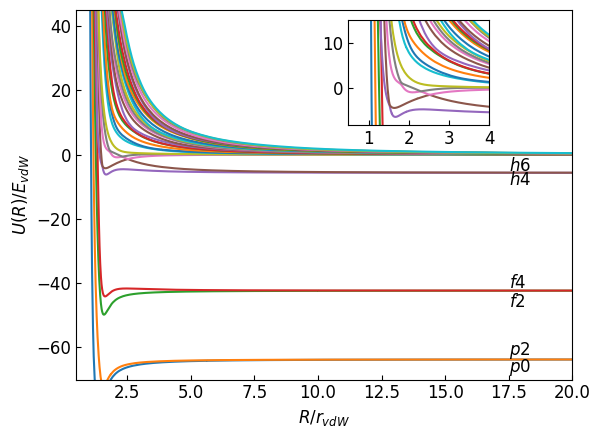}}
    \subfloat[\label{fig:sps2ndpwaveEig}]
    {\includegraphics[width=9cm]{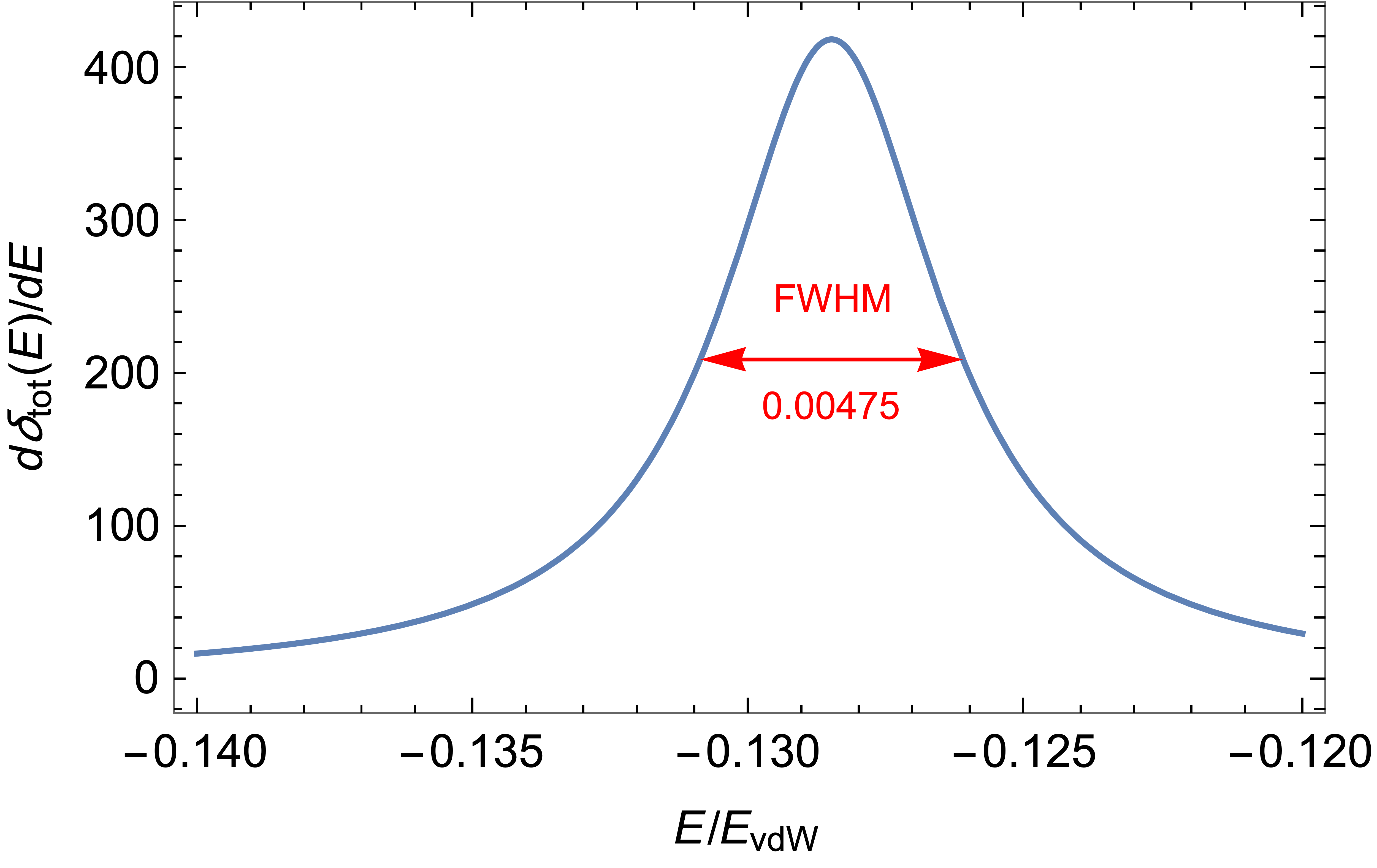}}
    \caption{(color online). (a) Three-body Born-Oppenheimer potential curves are presented for two spin-up and one spin-down fermion ($\uparrow\downarrow\uparrow$) with symmetry $L^{\Pi}=1^{-}$. The interaction between spin-up and spin-down fermions is set at the $s$-wave unitary limit with no deep dimers in different spin states, and the two spin-up fermionic interaction is set at the $2^{\text{nd}}$ $p$-wave pole, which has deep $p$-, $f$- and $h$-wave bound states. The letter labeling the potential curves converging to a negative energy represents the angular momentum quantum number $l$ of the dimer, and the number labels the relative angular momentum between the third atom and the dimer $l'$. The inset shows the detail of Born-Oppenheimer potential curves in the short range. (b)  Shown is the derivative of the sum of eigenphase shifts as a function of energy which is proportional to the time delay and corresponds to a resonance that is a universal $p$-wave trimer; in this case, and the resonance peak is located around $E=-0.129\, E_{\text{vdW}}$.
    }
    \label{fig:sps2ndpwave2}
\end{figure}

\twocolumngrid

However, if the absolute value of the $p$-wave scattering volume is increased beyond $\gtrsim -12\,r_{\text{vdW}}^{3}$, a universal trimer state can be discovered. As a result of the hyperradial potential curve barrier and depth of the lowest three-body continuum potential curve getting lower when $V_p$ gets larger and more negative (i.e., as the two-body interaction gets more attractive between the two spin-up fermions), the $p$-wave timer state does not have the restriction for the mass ratio $m_{\uparrow}/m_{\downarrow} > 8.172$ and it already exists for equal mass atoms. The $p$-wave unitary channel (the dashed line where $V_p
\rightarrow -\infty$) might initially appear to suggest the possibility of an Efimov effect (corresponding to an complex value of $l_{e,\nu}$ value corresponding to an attractive long-range potential curve with negative coefficient of $R^{-2}$). Indeed, if we consider the  Born-Oppenheimer potential curve which neglects the diagonal non-adiabatic coupling $Q_{\nu\nu}(R)$, there would be an Efimov effect with an infinity series of trimer states. However, the effective {\it adiabatic} potential curves that include this diagonal correction do not exhibit any evidence of the true Efimov effect, and this corrected $p$-wave unitary channel has an $l_e$ value that is very close to zero. 

\begin{figure}[htbp]
    \centering
    \subfloat[\label{fig:sps3rdpwave}]
    {\includegraphics[width=8.5cm]{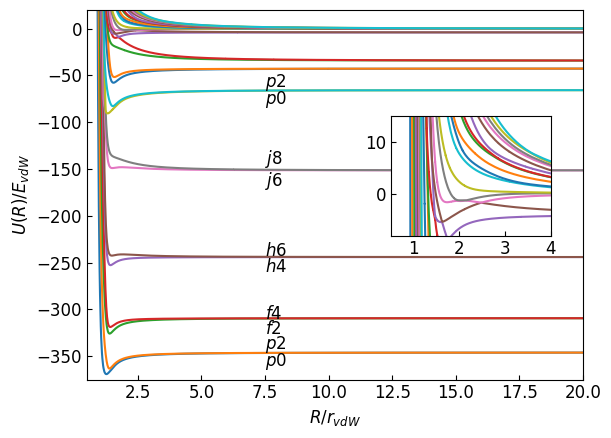}}
    \subfloat[\label{fig:sps3rdpwaveEig}]
    {\includegraphics[width=9cm]{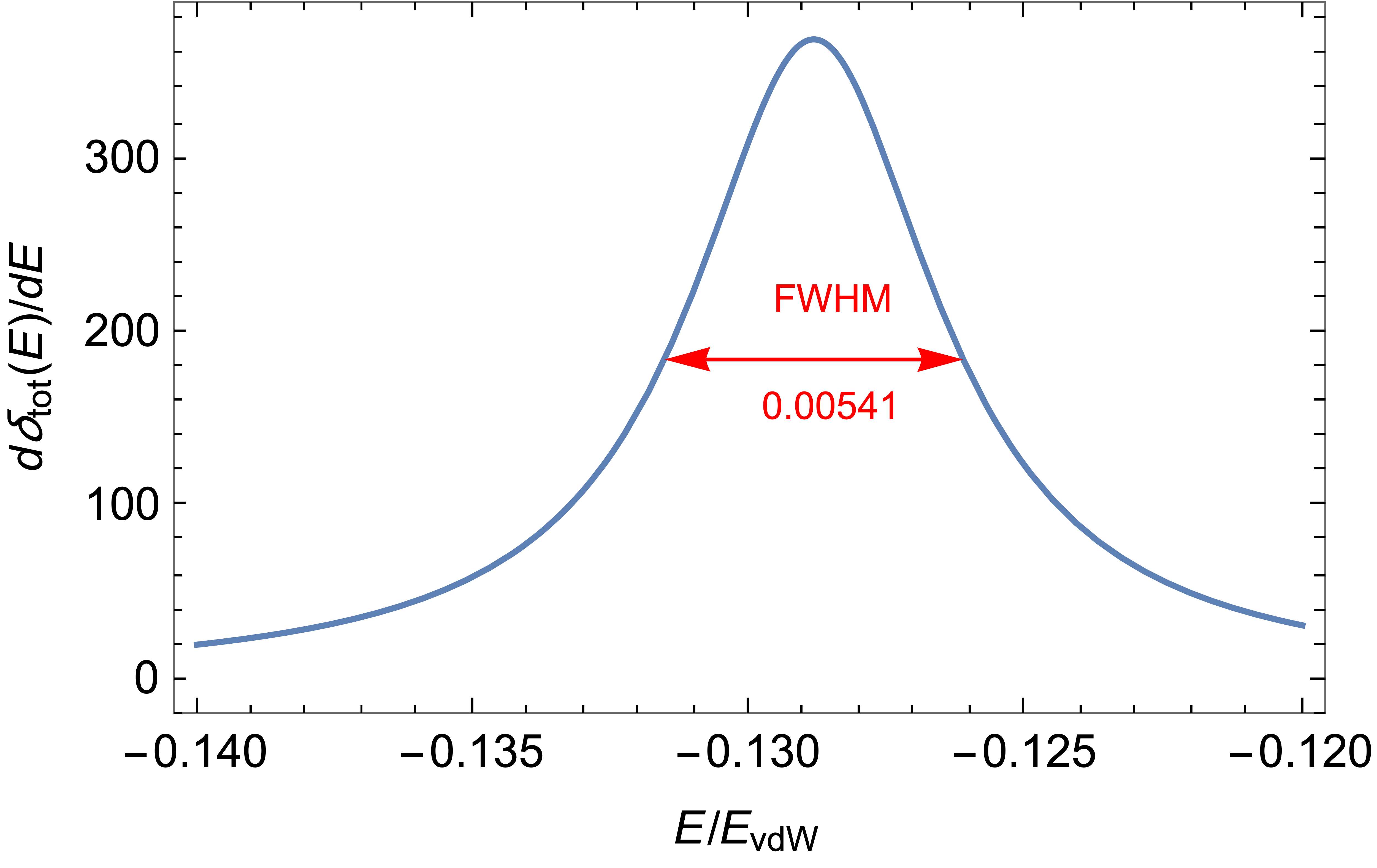}}
    \caption{(color online). (a) Three-body Born-Oppenheimer potential curves for two spin-up and one spin-down fermion ($\uparrow\downarrow\uparrow$) with total angular momentum and parity $L^{\Pi}=1^{-}$. The interaction between spin-up and spin-down fermions is set at the $s$-wave unitary limit with no deep opposite spin dimers, and the two spin-up fermion interaction is set at the $3^{\text{rd}}$ $p$-wave pole, which has deep $p$-, $f$-, $h$- and $j$-wave bound states. The letter represents the angular momentum quantum number $l$ of the dimer, and the number labels the angular momentum $l'$ between the third atom and the dimer. The inset shows expanded details of the Born-Oppenheimer potential curves in the short range. (b) The derivative of the sum of eigenphase shift is plotted as a function of energy near the universal $p$-wave trimer, and the peak is located around $E=-0.129\, E_{\text{vdW}}$.  The resonance decay width in van der Waals energy units is also indicated on the figure.}
    \label{fig:sps3rdpwave2}
\end{figure}

The behavior of the $p$-wave universal trimer is next explored in the range from the first $p$-wave pole to the second $p$-wave pole, still using a Hamiltonian based on two-body Lennard-Jones potentials. Fig.\ref{fig:sps2ndpwave} shows the Born-Oppenheimer potential curves for channels 1-30,keeping the interaction between unequal spin fermions fixed at the $1^{\text{st}}$ $s$-wave pole and keeping the interaction between the same spin fermions at the $2^{\text{nd}}$ $p$-wave pole. There are six atom-dimer channels with $p$-, $f$-, and $h$-wave bound states. Because there are six open channels, the sum of the eigenphase shifts (the eigenphase sum) is employed to analyze the $p$-wave universal trimer resonance energies. For the previous cases explored, there were no open channels so we could solve for the trimer bound states using a multi-channel calculation based on the slow-variable discretization (SVD) method\cite{tolstikhin1996slow}. The eigenphase shifts $\delta_i(E)$ can be obtained by diagonalizing the $\boldsymbol{\mathcal{K}}$-matrix\cite{wang2012hyperspherical, chen2022thesis} followed by taking the $\arctan$. Therefore, the total eigenphase shift can be written as\cite{gao1989energy,greene1987negative},
\begin{equation}\label{eq:eigenphase}
    \delta_{\text{tot}}(E)=\sum_{i=1}^{N_o}\delta_i(E)=\sum_{i=1}^{N_o}\tan^{-1}(\lambda_i),
\end{equation}
where $\lambda_i$ is the $i$-th eigenvalue of $\boldsymbol{\mathcal{K}}$-matrix, $E$ is the specified (collision) energy, and $N_o$ is the number of open channels. Fig.\ref{fig:sps2ndpwaveEig} illustrates the energy derivative of the eigenphase sum in Eq.(\ref{eq:eigenphase}), proportional to the total Wigner-Smith time-delay;  the resonance peak corresponds to the 3-body $p$-wave universal energy which is $E=-0.129\, E_{\text{vdW}}$. This energy is close to the previous result obtained with the interaction between the two spin-up fermions set at the $1^{\text{st}}$ $p$-wave pole, which is $E=-0.1355\, E_{\text{vdW}}$. Fig.\ref{fig:sps3rdpwave} plots the Born-Oppenheimer potential curves obtained when two spin-up fermions are set at the $3^{\text{rd}}$ $p$-wave pole and where the opposite spin fermion interaction is set at the first $s$-wave pole (i.e. infinite $a_s$). The inset shows the short-range behavior of these potential curves. There are 16 atom-dimer channels, corresponding to two each of spin-polarized dimer $p$-, $f$-, $h$- and $j$-wave bound states. The $p$-wave universal trimer can also be found in this situation, and the evidence is shown in Fig.\ref{fig:sps3rdpwaveEig}. The peak is located at around $E=-0.129\, E_{\text{vdW}}$ which is similar to the universal trimer energies for two spin-up fermions interacting at the $1^{\text{st}}$ and $2^{\text{nd}}$ $p$-wave poles. 

\onecolumngrid

\twocolumngrid

\onecolumngrid

\begin{table}[ht]
    \begin{tabular}{ccccc}
        \hline\hline
        $L^{\Pi}$ & \rule{50pt}{0pt}Trimer Exists? (w/o $Q_{\nu\nu}(R)$) & \rule{50pt}{0pt} Trimer Exists? (w/ $Q_{\nu\nu}(R)$) & \rule{50pt}{0pt} Universality of Trimer? \\
        \hline
        $0^{+}$ & \rule{50pt}{0pt} no  & \rule{50pt}{0pt} no & \rule{50pt}{0pt} no & \\
        $1^{+}$ & \rule{50pt}{0pt} yes & \rule{50pt}{0pt} no & \rule{50pt}{0pt} no & \\
        $1^{-}$ & \rule{50pt}{0pt} faux-Efimov & \rule{50pt}{0pt} yes & \rule{50pt}{0pt} yes & \\
        $2^{-}$ & \rule{50pt}{0pt} no  & \rule{50pt}{0pt} no & \rule{50pt}{0pt} no & \\
        \hline\hline
    \end{tabular}
    \caption{Comparison of the $p$-wave universal trimer status for different symmetries $L^\Pi$, with the interaction between unequal spin fermions $(\uparrow\downarrow)$ fixed at the $s$-wave unitary limit and the interaction between the two spin-up fermion $(\uparrow\uparrow)$ set at the $p$-wave unitary limit. The ``faux-Efimov'' effect is present in the lowest Born-Oppenheimer potential curve and has a negative coefficient of $1/R^2$ asymptotically. The ``Universality of Trimer'' column means that we have tested the different $s$- and $p$-wave poles for each pair of fermion assess the possibility that the bound trimer energy is in fact universal.}
    \label{tab:spsuniversal}
\end{table}

\twocolumngrid

Table \ref{tab:spsuniversal} shows the $p$-wave universal trimer state possibilities that have been observed for different symmetries $L^\Pi$ in equal mass systems.  In some cases, $p$-wave universal trimers would appear to occur in our calculations when the diagonal adiabatic correction (also known as the Born-Huang correction) $Q_{\nu\nu}(R)$ matrix element is omitted; in particular this occurs for trimer angular momentum $L^\Pi=1^+$ and $1^-$. We have designated such cases as a ``faux-Efimov'' effect which means that it only exists in the approximation where the diagonal correction $Q_{\nu\nu}(R)$ matrix element is omitted from the effective adiabatic potential curve.  Specifically, the Born-Oppenheimer potential curve in that case has a negative coefficient of $R^{-2}$ and a complex value of $l_e$ for the symmetry $L^\Pi=1^-$. However, this case is not a {\it true} Efimov effect because the $1/R^2$ coefficient of the {\it full} three-body effective adiabatic potential curve does not have any attraction at large $R$ [see Ref\cite{chen2022efimov} Fig.7]. The $p$-wave universal trimer state identified here has symmetry $L^\Pi=1^-$, and its universality is tested by considering two-body interactions at different $s$-wave poles for the unequal spin fermions and different $p$-wave poles for the same spin fermions. Table \ref{tab:sps} shows the comparison of the $l_{e,\nu}$ value obtained when  the $s$-wave or $p$-wave interaction is tuned to the unitary limit for unequal spin fermions and/or at the $p$-wave unitarity limit between the two spin-up fermions with different trimer angular momentum. If the spin-up and spin-down fermions interact at $s$-wave unitarity, the $l_{e,\nu}$ value becomes modified at large hyperradii to an irrational number, which can be calculated by solving a transcendental equation\cite{blume2007universal,werner2006unitary}, and these new values (bold number) also change the $l_{e,\nu}$ value of one of the degenerate states. However, the $p$-wave type of the unitary channel (underlined and integer number) not modifies the degenerate state of $l_{e,\nu}$. It changes in the lowest potential curve, and its $l_{e,\nu}$ values are an integer. For example, in the third row of Table \ref{tab:sps}, the $l_{e,\nu}$ values for the symmetry $L^{\Pi}=1^-$ change from half-integer to integer for both the lowest and second-lowest values. In contrast, for the symmetries $L^{\Pi}=0^+$, $1^+$, and $2^-$, there is only one lowest potential curve each, where the $l_{e,\nu}$ values also transition from half-integer or irrational numbers to integers.

It should be kept in mind that there are no $s$-wave interactions between spin-up and spin-down fermionic atoms that are present for symmetries $L^\Pi=1^+$ and $2^-$.
Hence, going to the $s$-wave unitary limit for the interaction between unequal spin fermions does not change the coefficient of $1/R^2$ asymptotically for those two symmetries. 

\onecolumngrid

\begin{table}[htbp]
    \centering    
    \begin{tabular}{c|cccccccccc}
        \hline\hline 
        $ L^\Pi$ &&&&& \rule{20pt}{0pt} $l_{e,\nu}$  &&&&\\
        \hline
        \multirow{8}{*}{$0^+$}  & \rule{20pt}{0pt} 7/2 & \rule{20pt}{0pt} 11/2 & \rule{20pt}{0pt} 15/2 & \rule{20pt}{0pt} 15/2 & \rule{20pt}{0pt} 19/2 & \rule{20pt}{0pt} 19/2 & \rule{20pt}{0pt} 23/2 & \rule{20pt}{0pt} 23/2 & \rule{20pt}{0pt} 23/2   \\
        &&&&\multicolumn{3}{c}{ \rule{20pt}{0pt}  non-interacting}&&& \\
        & \rule{20pt}{0pt} \textbf{1.666} & \rule{20pt}{0pt} \textbf{4.627} & \rule{20pt}{0pt} \textbf{6.614} & \rule{20pt}{0pt} 15/2 & \rule{20pt}{0pt} \textbf{8.332} & \rule{20pt}{0pt} 19/2 & \rule{20pt}{0pt} \textbf{10.562} & \rule{20pt}{0pt} 23/2 & \rule{20pt}{0pt} 23/2 \\
        &&&\multicolumn{5}{c}{\rule{20pt}{0pt} $\uparrow\downarrow$ at $s$-wave unitary limit and $\uparrow\uparrow$ non-interacting} &&\\ & \rule{20pt}{0pt} \underline{1} & \rule{20pt}{0pt} 7/2 & \rule{20pt}{0pt} 11/2 & \rule{20pt}{0pt} 15/2 & \rule{20pt}{0pt} 15/2 & \rule{20pt}{0pt} 19/2 & \rule{20pt}{0pt} 19/2 & \rule{20pt}{0pt} 23/2 & \rule{20pt}{0pt} 23/2 \\
        &&&\multicolumn{5}{c}{\rule{20pt}{0pt} $\uparrow\uparrow$ at $p$-wave unitary limit and $\uparrow\downarrow$ non-interacting} &&\\
        & \rule{20pt}{0pt} \underline{1} & \rule{20pt}{0pt} \textbf{1.666} & \rule{20pt}{0pt} \textbf{4.627} & \rule{20pt}{0pt} \textbf{6.614} & \rule{20pt}{0pt} 15/2 & \rule{20pt}{0pt} \textbf{8.332} & \rule{20pt}{0pt} 19/2 & \rule{20pt}{0pt} \textbf{10.562} & \rule{20pt}{0pt} 23/2 \\
        &&\multicolumn{7}{c}{\rule{20pt}{0pt} $\uparrow\downarrow$ at $s$-wave unitarity and $\uparrow\uparrow$ at $p$-wave unitary limit}\\
        \hline
        \multirow{8}{*}{$1^+$}  & \rule{20pt}{0pt} 7/2 & \rule{20pt}{0pt} 11/2 & \rule{20pt}{0pt} 15/2 & \rule{20pt}{0pt} 15/2 & \rule{20pt}{0pt} 19/2 & \rule{20pt}{0pt} 19/2 & \rule{20pt}{0pt} 23/2 & \rule{20pt}{0pt} 23/2 & \rule{20pt}{0pt} 23/2   \\
        &&&&\multicolumn{3}{c}{\rule{20pt}{0pt} non-interacting}&&& \\
        & \rule{20pt}{0pt} 7/2 & \rule{20pt}{0pt} 11/2 & \rule{20pt}{0pt} 15/2 & \rule{20pt}{0pt} 15/2 & \rule{20pt}{0pt} 19/2 & \rule{20pt}{0pt} 19/2 & \rule{20pt}{0pt} 23/2 & \rule{20pt}{0pt} 23/2 & \rule{20pt}{0pt} 23/2 \\
        &&&\multicolumn{5}{c}{\rule{20pt}{0pt} $\uparrow\downarrow$ at $s$-wave unitary limit and $\uparrow\uparrow$ non-interacting}&&\\ & \rule{20pt}{0pt} \underline{1} & \rule{20pt}{0pt} 7/2 & \rule{20pt}{0pt} 11/2 & \rule{20pt}{0pt} 15/2 & \rule{20pt}{0pt} 15/2 & \rule{20pt}{0pt} 19/2 & \rule{20pt}{0pt} 19/2 & \rule{20pt}{0pt} 23/2 & \rule{20pt}{0pt} 23/2 \\
        &&&\multicolumn{5}{c}{$\uparrow\uparrow$ at $p$-wave unitary limit and $\uparrow\downarrow$ non-interacting}&&\\
        & \rule{20pt}{0pt} \underline{1} & \rule{20pt}{0pt} 7/2 & \rule{20pt}{0pt} 11/2 & \rule{20pt}{0pt} 15/2 & \rule{20pt}{0pt} 15/2 & \rule{20pt}{0pt} 19/2 & \rule{20pt}{0pt} 19/2 & \rule{20pt}{0pt} 23/2 & \rule{20pt}{0pt} 23/2  \\
        &&\multicolumn{7}{c}{\rule{20pt}{0pt} $\uparrow\downarrow$ at $s$-wave unitarity and $\uparrow\uparrow$ at $p$-wave unitary limit}\\
        \hline
        \multirow{8}{*}{$1^-$}  & \rule{20pt}{0pt} 5/2 & \rule{20pt}{0pt} 9/2 & \rule{20pt}{0pt} 9/2 & \rule{20pt}{0pt} 13/2  & \rule{20pt}{0pt} 13/2 & \rule{20pt}{0pt} 13/2 & \rule{20pt}{0pt} 17/2 & \rule{20pt}{0pt} 17/2 & \rule{20pt}{0pt} 17/2  \\
        &&&&\multicolumn{3}{c}{\rule{20pt}{0pt} non-interacting}&& \\
        &  \rule{20pt}{0pt} \textbf{1.272} & \rule{20pt}{0pt} \textbf{3.858} & \rule{20pt}{0pt} 9/2 & \rule{20pt}{0pt} \textbf{5.216} & \rule{20pt}{0pt} 13/2 & \rule{20pt}{0pt} 13/2 & \rule{20pt}{0pt} \textbf{7.553} & \rule{20pt}{0pt} 17/2 & \rule{20pt}{0pt} 17/2 \\
        &&&\multicolumn{5}{c}{\rule{20pt}{0pt} $\uparrow\downarrow$ at $s$-wave unitary limit and $\uparrow\uparrow$ non-interacting}&&\\
        & \rule{20pt}{0pt} \underline{0} & \rule{20pt}{0pt} \underline{2} & \rule{20pt}{0pt} 5/2 & \rule{20pt}{0pt} 9/2 & \rule{20pt}{0pt} 9/2 & \rule{20pt}{0pt} 13/2  & \rule{20pt}{0pt} 13/2 & \rule{20pt}{0pt} 13/2 & \rule{20pt}{0pt} 17/2 \\
        &&&\multicolumn{5}{c}{\rule{20pt}{0pt} $\uparrow\uparrow$ at $p$-wave unitary limit and $\uparrow\downarrow$ non-interacting}&&\\
        & \rule{20pt}{0pt} \underline{0} & \rule{20pt}{0pt} \textbf{1.272} & \rule{20pt}{0pt} \underline{2} & \rule{20pt}{0pt} \textbf{3.858} & \rule{20pt}{0pt} 9/2 & \rule{20pt}{0pt} \textbf{5.216} & \rule{20pt}{0pt} 13/2 & \rule{20pt}{0pt} 13/2 & \rule{20pt}{0pt} \textbf{7.553} &\\
        &&\multicolumn{7}{c}{\rule{20pt}{0pt} $\uparrow\downarrow$ at $s$-wave unitarity and $\uparrow\uparrow$ at $p$-wave unitary limit}\\
        \hline
        \multirow{8}{*}{$2^-$}  & \rule{20pt}{0pt} 9/2 & \rule{20pt}{0pt} 9/2 & \rule{20pt}{0pt} 13/2 & \rule{20pt}{0pt} 13/2 & \rule{20pt}{0pt} 13/2 & \rule{20pt}{0pt} 17/2 & \rule{20pt}{0pt} 17/2 & \rule{20pt}{0pt} 17/2 & \rule{20pt}{0pt} 21/2  \\
        &&&&\multicolumn{3}{c}{\rule{20pt}{0pt} non-interacting}&&& \\
        & \rule{20pt}{0pt} 9/2 & \rule{20pt}{0pt} 9/2 & \rule{20pt}{0pt} 13/2 & \rule{20pt}{0pt} 13/2 & \rule{20pt}{0pt} 13/2 & \rule{20pt}{0pt} 17/2 & \rule{20pt}{0pt} 17/2 & \rule{20pt}{0pt} 17/2 & \rule{20pt}{0pt} 21/2 \\
        &&&\multicolumn{5}{c}{\rule{20pt}{0pt} $\uparrow\downarrow$ at $s$-wave unitary limit and $\uparrow\uparrow$ non-interacting}&&\\
        & \rule{20pt}{0pt} \underline{2} & \rule{20pt}{0pt} 9/2 & \rule{20pt}{0pt} 9/2 & \rule{20pt}{0pt} 13/2 & \rule{20pt}{0pt} 13/2 & \rule{20pt}{0pt} 13/2 & \rule{20pt}{0pt} 17/2 & \rule{20pt}{0pt} 17/2 & \rule{20pt}{0pt} 17/2  \\
        &&&\multicolumn{5}{c}{\rule{20pt}{0pt} $\uparrow\uparrow$ at $p$-wave unitary limit and $\uparrow\downarrow$ non-interacting}&&\\
        & \rule{20pt}{0pt} \underline{2} & \rule{20pt}{0pt} 9/2 & \rule{20pt}{0pt} 9/2 & \rule{20pt}{0pt} 13/2 & \rule{20pt}{0pt} 13/2 & \rule{20pt}{0pt} 13/2 & \rule{20pt}{0pt} 17/2 & \rule{20pt}{0pt} 17/2 & \rule{20pt}{0pt} 17/2  \\
        &&\multicolumn{7}{c}{\rule{20pt}{0pt} $\uparrow\downarrow$ at $s$-wave unitarity and $\uparrow\uparrow$ at $p$-wave unitary limit}\\
        \hline\hline
    \end{tabular}
    \caption{Comparison of the $1^{\text{st}}$ to $9^{\text{th}}$ $l_{e,\nu}$ values for trimers consisting of two spin-up and one spin-down fermion $(\downarrow\uparrow\uparrow)$ with the non-interacting values of $l_{e,\nu}$. Cases considered involve different spin fermions that are either at $s$-wave unitarity or else are noninteracting as are labeled in the Table.  Other cases show the same spin fermions either at $p$-wave unitarity or noninteracting. The bold and underlined numbers correspond to the $l_{e,\nu}$ values of the $s$-wave unitary channel and the $p$-wave unitary channel, respectively.  
    }
    \label{tab:sps}
\end{table}

\twocolumngrid

\section{Two-component fermionic trimers at $p$-wave unitarity}\label{sec:p-wave}

This section treats two types of $p$-wave universal trimers occurring for different symmetries, and explores the extent of their universality. The first case has the interaction between the different spin fermions set close to $p$-wave unitarity while that between two spin-up fermions is a comparatively weak interaction. The second case considers the $p$-wave universal interaction between each pair of fermions with various symmetries.

\subsection{Opposite fermions only interact close to $p$-wave unitarity}\label{subsec:p-wave_only_opposite}

\begin{figure}[htbp]
    \includegraphics[width=8.5cm]{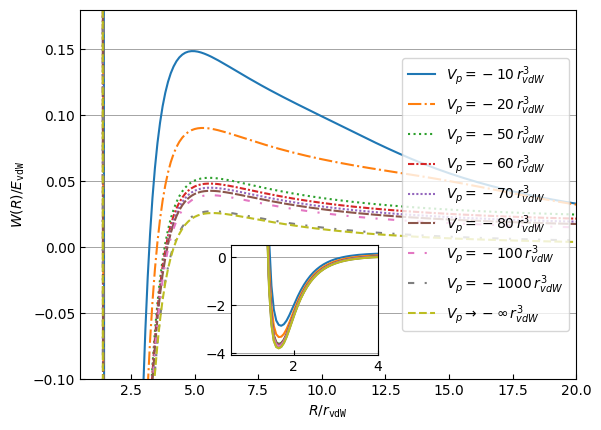}
    \caption{(color online). The lowest effective adiabatic potential curves are shown versus the hyperradius for several different $p$-wave scattering volumes ($V_p$) in van der Waals units of length and energy, for the system $\downarrow\uparrow\uparrow$ that has weak interaction between the two spin-up fermions for symmetry $L^{\Pi}=1^+$. As $\lvert V_p \rvert$ gets larger, the potential curve approaches the three-body threshold with a centrifugal barrier strength close to the value implied by the reduced value $l_e=1$ in the adiabatic potential energy curve. The inset shows the evolution of the depth of the effective adiabatic potential curves near their minima.}
    \label{fig:pspVp}
\end{figure}

In this subsection, two cases are discussed, where the $p$-wave two-body interaction is close to unitarity between either different spin fermions or between two fermions in the same spin state. The situation where the different spin fermions interact at the $p$-wave unitarity limit and with weak interaction between the two spin-up fermions has been mentioned in previous work by Ref.\cite{braaten2012renormalization} where those authors originally claimed that there would be a $p$-wave Efimov effect; however, in the hyperspherical viewpoint, there is a faux-Efimov effect, as was described above, where the infinite series of Efimov states {\it appears} (incorrectly) to occur at $p$-wave unitarity if one does neglects the diagonal $Q_{\nu\nu}(R)$, for trimer angular momentum $L^\Pi=1^-$. Additionally, it is important to note that opposite spin fermions close to the $p$-wave unitary limit still exhibit an $s$-wave interaction, with the $s$-wave scattering length around $a_s \approx 1.988$ $r_\text{vdW}$. The three-body Hamiltonian does still include $s$-wave interactions even though they are far from unitarity. We stress that it is important to included this diagonal correction in general because it is part of the Hamiltonian. But, as previously mentioned in Sec.III B.2(in Ref.\cite{chen2022efimov}), there is no $p$-wave Efimov effect for symmetries $L^\Pi=0^+$, $1^+$ and $1^-$, although we have found a single $p$-wave universal trimer for symmetry $L^\Pi=1^+$ in the presence of van der Waals interactions. Fig.\ref{fig:pspVp} plots the lowest three-body effective adiabatic potential curve as a function of the $p$-wave scattering volume $V_p$ between spin-up and spin-down fermion for the trimer symmetry $L^\Pi=1^+$. The interaction between two spin-up fermionic atoms is fixed at $V_p \approx -2\,r_{\text{vdW}}^3$. A universal three-body bound state can be formed when the $\lvert V_p \rvert$ grows very large in magnitude, as the hyperradial potential barrier decreases and the depth of the relevant effective adiabatic potential curve gets deeper.

Table \ref{tab:pspuniversal} illustrates the $p$-wave universal trimer status with respect to different symmetries $L^\Pi$. A $p$-wave universal trimer is predicted to exist with trimer angular momentum $L^\Pi=1^+$, and its universality has been tested for various $p$-wave poles of the two-body Lennard-Jones potential. There is a ``faux-Efimov'' effect which has a negative $1/R^2$ coefficient in the Born-Oppenheimer potential curve which neglects the diagonal correction term. However, in the more relevant effective {\it adiabatic} potential curves that include the diagonal correction, there is no actual Efimov effect and no $p$-wave universal trimer with symmetry $L^\Pi=1^-$.  Fig.\ref{fig:ppptrimer} shows $p$-wave universal trimer state energies as functions of the $p$-wave scattering length $(a_p)$, which were calculated by including 30 coupled continuum channel potential curves. Here the $p$-wave scattering volume (or length) $V_p$ (or $a_p$) and $V'_p$ (or $a'_p$) represent the interaction between opposite spin fermions and two spin-up fermions, respectively. Different symbols represent the different interaction situations for each pair of fermions or different spin states of three fermionic atom. Squares (filled gold) are fixed at $V'_p=-2\, r_{\text{vdW}}^3$ and we tune the value of  $V_p$ which the universal trimer energy is around $E\approx -0.624\, E_{\text{vdW}}$. The other two curves (diamonds and triangles) have the same universal trimer energies, which is $E\approx -2.0216\, E_{\text{vdW}}$. Diamonds (solid green) are set the two spin-up fermion close to divergence $V'_p=\infty \, r_{\text{vdW}}^3$ and change the interaction between unequal spin fermion $V_p$. Triangles (solid red) show the fixed interaction within unequal spin fermion at $p$-wave unitary limit $(V'_p=\infty \, r_{\text{vdW}}^3)$ and tune the $V'_p$ between equal spin fermion. The diamond points have larger $|V_p|$ than triangle points at the zero trimer energy because there is just one pair of fermion settings at the $p$-wave unitary limit. In the calculation represented by triangles, there are two pairs of Fermi gases fixed at the $p$-wave pole (unitarity). Therefore, in the diamond case, the starting point at which the universal trimer state exists needs to have a large $p$-wave scattering volume (length).

\onecolumngrid

\begin{table}[htbp]
    \centering
    \begin{tabular}{ccccc}
    \hline\hline
    \rule{0pt}{10pt} $L^{\Pi}$ & \rule{50pt}{0pt} Trimer Exists? (w/o $Q_{\nu\nu}(R)$) & \rule{50pt}{0pt} Trimer Exists? (w/ $Q_{\nu\nu}(R)$) & \rule{50pt}{0pt} Universality of Trimer? \\
    \hline
    $0^{+}$ & \rule{50pt}{0pt} no  & \rule{50pt}{0pt} no & \rule{50pt}{0pt} no & \\
    $1^{+}$ & \rule{50pt}{0pt} yes & \rule{50pt}{0pt} yes & \rule{50pt}{0pt} yes & \\
    $1^{-}$ & \rule{50pt}{0pt} faux-Efimov & \rule{50pt}{0pt} no & \rule{50pt}{0pt} no & \\
    $2^{-}$ & \rule{50pt}{0pt} no & \rule{50pt}{0pt} no & \rule{50pt}{0pt} no & \\
    \hline\hline
    \end{tabular}
    \caption{Comparison of $p$-wave universal trimer status, i.e. whether a trimer state is universal) for different symmetries $L^\Pi$, with the interaction between opposite spin fermions $(\uparrow\downarrow)$ set at the $p$-wave unitary limit while the two spin-up fermion $(\uparrow\uparrow)$ have weak interaction. The ``faux-Efimov'' case is represented by the lowest Born-Oppenheimer potential curve and it has a negative coefficient of $1/R^2$ asymptotically. The ``Universality of Trimer'' column  tests whether the universal trimer energy occurs consistently for different $p$-wave poles for the interaction between different spin state fermions.}
    \label{tab:pspuniversal}
\end{table}

\twocolumngrid

\subsection{Opposite and same spin fermions are all interacting at the $p$-wave unitary limit.}\label{subsec:p-wave_all}

\begin{figure}[htbp]
    \centering
    \subfloat[\label{fig:ppptrimer}]
    {\includegraphics[width=9cm]{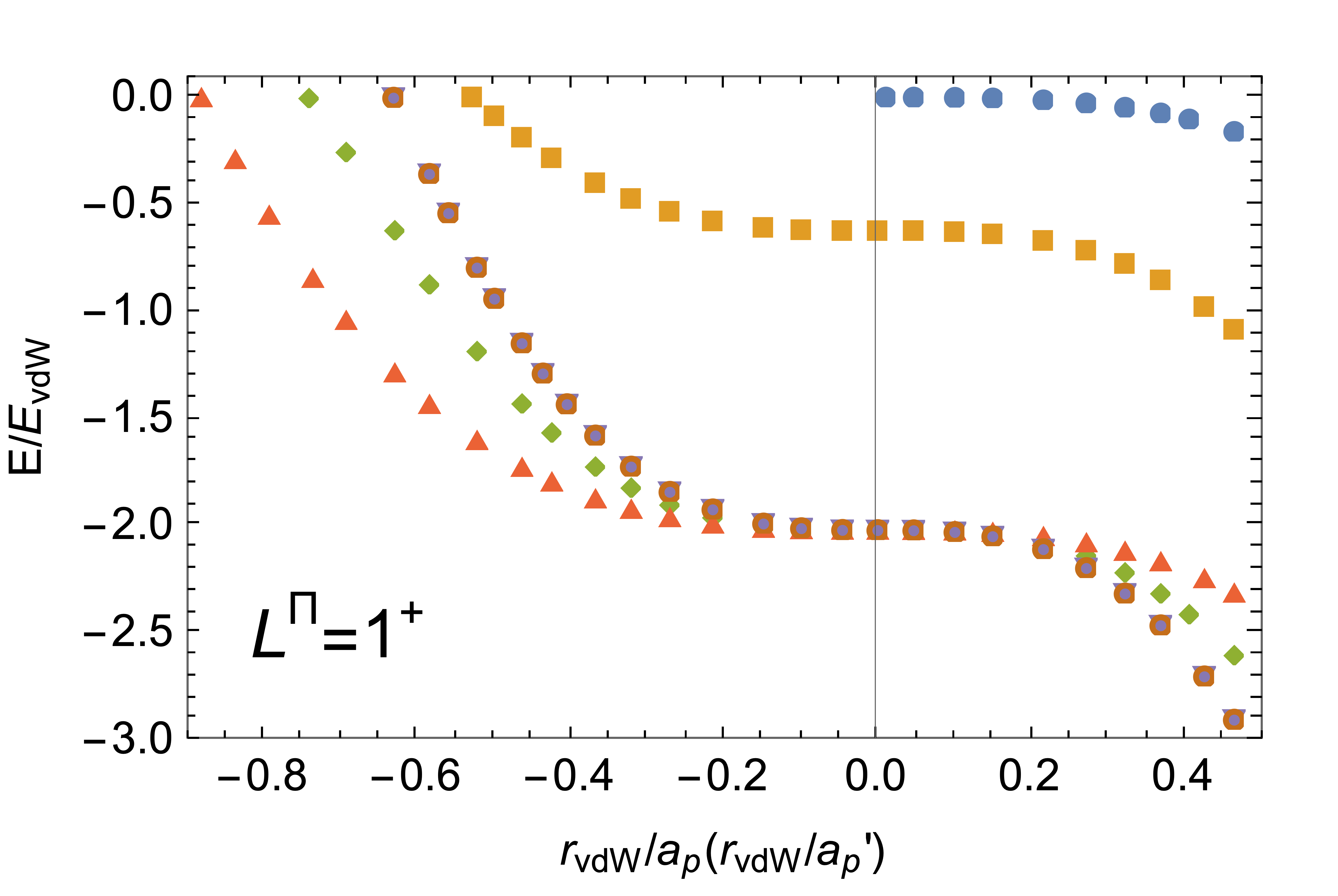}}
    \subfloat[\label{fig:pppJ1mtrimer}]
    {\includegraphics[width=9cm]{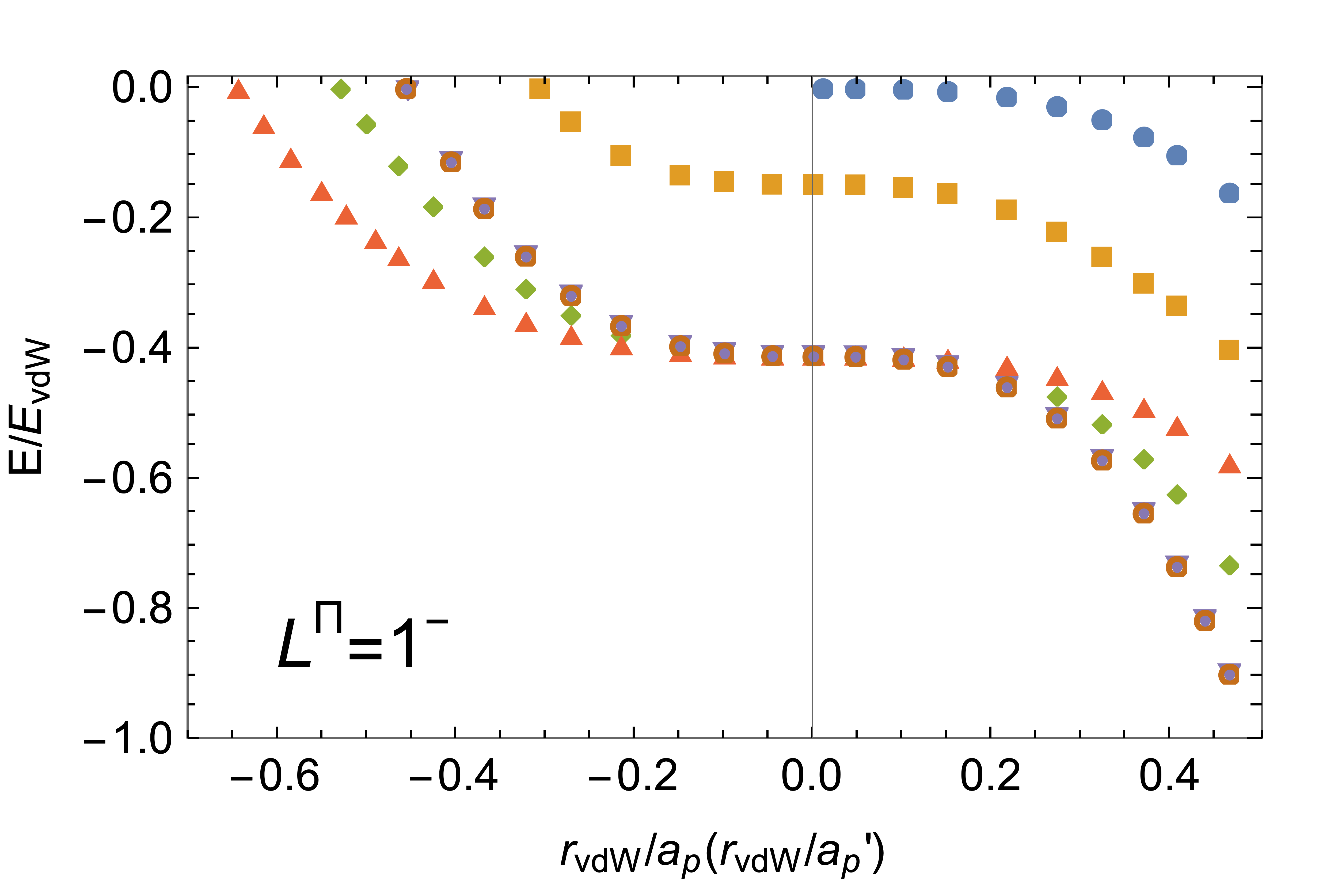}}
    \caption{(color online). Shown are $p$-wave universal trimer state energies for the system of one spin-down and two spin-up fermions ($\uparrow\downarrow\uparrow$), plotted versus the inverse of the $p$-wave scattering length  ($a_p \equiv V_{p}^{1/3}$). The $V_p(a_p)$ and $V'_p(a'_p)$ represent respectively the $p$-wave scattering volume (or length) between unequal spin fermion and same spin fermions. Circles (solid blue) show the two-body $p$-wave bound state. Squares (solid gold) are for fixed $V'_p=-2\, r_{\text{vdW}}^{3}$ and with $V_p$ tuned. Inverted-triangles are with $V'_p=V_p$ being tuned. Diamonds (solid green) are for a fixed $V'_p=\infty\, r_{\text{vdW}}^{3}$ and with $V_p$ tuned. Triangles (solid red) are for fixed $V_p=\infty\, r_{\text{vdW}}^{3}$ and with $V'_p$ tuned. The open (red) circles are for three spin-up fermion and all interactions $V_p$ tuned simultaneously. (a), trimer symmetry $L^{\Pi}=1^+$. (b), trimer symmetry $L^{\Pi}=1^-$}
    \label{fig:pppJ1pJ1mtrimer}
\end{figure}

In Fig.\ref{fig:ppptrimer}, the inverted-triangles plot of universal trimer energies is shown as a function of the $p$-wave scattering length, and its energy at the original $(a_p\rightarrow \infty)$ is close to $E\approx -2.0216\, E_{\text{vdW}}$. The universal trimer energies of three spin-up fermion at $p$-wave unitary limit (open circles) are slightly higher than the two spin-up and one spin-down Fermi gases at $p$-wave unitarity. The interpretation of this result is discussed in the next section. Table \ref{tab:pppuniversal} shows the universal trimer state and with and without diagonal $Q_{\nu\nu}(R)$ correction with respect to different symmetries $L^\Pi$. As previous mention, the ``faux-Efimov'' means that the coefficient of the Born-Oppenheimer potential curves has a negative value (purely attractive) at spatial angular momentum $L^\Pi=1^+$. A more interesting result we have not observed previously is that there is a ``degenerate faux-Efimov'' case in the symmetry $L^\Pi=1^-$, namely that when both the unequal spin fermions and the two spin-up fermion, i.e. all pairwise interactions, are set at $p$-wave unitarity, this produces an asymptotic degeneracy in the lowest Born-Oppenheimer potential curves, i.e. degenerate only at $R\rightarrow \infty$. 
In this case, the universal trimer exists at the two different trimer angular momentum $L^\Pi=1^+$ and $1^-$, and the trimer state also appears with different $p$-wave poles. In Table \ref{tab:psp}, the $p$-wave unitary channel is found (representing the underlined number) while the $p$-wave interaction of each pair of fermion either opposite spin or spin-polarized goes to infinity, and the interactions between every pair of fermion is set at the $p$-wave unitary limit. The asymptotically doubly-degenerate state of $p$-wave unitary channels also can be found, and its states are constructed from each pair of opposite as well as same spin Fermi gases. The reason for the degeneracy of the state can be seen as the orbital momentum of the pair of unequal spin fermion relative to the spin-up fermion has the $l_e=1$; on the other hand, the angular momentum between the pair of equal spin fermion and the spin-down fermion is $l_e=1$ for symmetry $L^\Pi=1^+$. Similar reasoning applies to the case of trimer symmetry $L^\Pi=1^-$, which also shows a doubly-degenerate state (asymptotically) of $p$-wave unitary channels.

\onecolumngrid

\twocolumngrid

\onecolumngrid

\begin{table}[htbp]
    \centering
    \begin{tabular}{ccccc}
    \hline\hline
    \rule{0pt}{10pt} $L^{\Pi}$ & \rule{50pt}{0pt} Trimer Exists? (w/o $Q_{\nu\nu}(R)$) & \rule{50pt}{0pt} Trimer Exists? (w/ $Q_{\nu\nu}(R)$) & \rule{50pt}{0pt} Universality of Trimer? \\
    \hline
    $0^{+}$ & \rule{50pt}{0pt} no  & \rule{50pt}{0pt} no & \rule{50pt}{0pt} no & \\
    $1^{+}$ & \rule{50pt}{0pt} yes & \rule{50pt}{0pt} yes & \rule{50pt}{0pt} yes & \\
    $1^{-}$ & \rule{50pt}{0pt} two faux-Efimov potentials & \rule{50pt}{0pt} yes & \rule{50pt}{0pt} yes & \\
    $2^{-}$ & \rule{50pt}{0pt} no & \rule{50pt}{0pt} no & \rule{50pt}{0pt} no & \\
    \hline\hline
    \end{tabular}
    \caption{Comparison of $p$-wave trimer universality status for different symmetries $L^\Pi$ with the interactions between equal and unequal spin state fermion all set at the $p$-wave unitary limit. The ``faux-Efimov'' case occurs when the lowest Born-Oppenheimer potential curve converging to the three-body threshold at 0 energy has a coefficient of $1/R^2$  that is more negative asymptotically than $-1/8R^2$ in van der Waals units. The ``Universality of Trimer'' column results are determined by testing the robustness of the trimer energy for different $p$-wave poles.
    }
    \label{tab:pppuniversal}
\end{table}

\begin{table}[htbp]
    \centering
    \begin{tabular}{c|cccccccccc}
    \hline\hline 
        \rule{0pt}{10pt} $L^\Pi$ &&&&& \rule{20pt}{0pt} $l_{e,\nu}$  &&&&\\
        \hline
        \multirow{8}{*}{$0^+$}  & \rule{20pt}{0pt} 7/2 & \rule{20pt}{0pt} 11/2 & \rule{20pt}{0pt} 15/2 & \rule{20pt}{0pt} 15/2 & \rule{20pt}{0pt} 19/2 & \rule{20pt}{0pt} 19/2 & \rule{20pt}{0pt} 23/2 & \rule{20pt}{0pt} 23/2 & \rule{20pt}{0pt} 23/2   \\
        &&&&\multicolumn{3}{c}{ \rule{20pt}{0pt} non-interacting}&&& \\
        & \rule{20pt}{0pt} \underline{1} & \rule{20pt}{0pt} 7/2 & \rule{20pt}{0pt} 11/2 & \rule{20pt}{0pt} 15/2 & \rule{20pt}{0pt} 15/2 & \rule{20pt}{0pt} 19/2 & \rule{20pt}{0pt} 19/2 & \rule{20pt}{0pt} 23/2 & \rule{20pt}{0pt} 23/2 \\
        &&&\multicolumn{5}{c}{\rule{20pt}{0pt} $\uparrow\downarrow$ at $p$-wave unitary limit and $\uparrow\uparrow$ non-interacting}&&\\ & \rule{20pt}{0pt} \underline{1} & \rule{20pt}{0pt} 7/2 & \rule{20pt}{0pt} 11/2 & \rule{20pt}{0pt} 15/2 & \rule{20pt}{0pt} 15/2 & \rule{20pt}{0pt} 19/2 & \rule{20pt}{0pt} 19/2 & \rule{20pt}{0pt} 23/2 & \rule{20pt}{0pt} 23/2 \\
        &&&\multicolumn{5}{c}{\rule{20pt}{0pt} $\uparrow\uparrow$ at $p$-wave unitary limit and $\uparrow\downarrow$ non-interacting}&&\\
        & \rule{20pt}{0pt} \underline{1} & \rule{20pt}{0pt} \underline{1} & \rule{20pt}{0pt} 7/2 & \rule{20pt}{0pt} 11/2 & \rule{20pt}{0pt} 15/2 & \rule{20pt}{0pt} 15/2 & \rule{20pt}{0pt} 19/2 & \rule{20pt}{0pt} 19/2 & \rule{20pt}{0pt} 23/2 \\
        &&&\multicolumn{5}{c}{\rule{20pt}{0pt} $\uparrow\downarrow\uparrow$ all at $p$-wave unitary limit}&&\\
        \hline
        \multirow{8}{*}{$1^+$}  & \rule{20pt}{0pt} 7/2 & \rule{20pt}{0pt} 11/2 & \rule{20pt}{0pt} 15/2 & \rule{20pt}{0pt} 15/2 & \rule{20pt}{0pt} 19/2 & \rule{20pt}{0pt} 19/2 & \rule{20pt}{0pt} 23/2 & \rule{20pt}{0pt} 23/2 & \rule{20pt}{0pt} 23/2   \\
        &&&&\multicolumn{3}{c}{\rule{20pt}{0pt} non-interacting}&&& \\
        & \rule{20pt}{0pt} \underline{1} & \rule{20pt}{0pt} 7/2 & \rule{20pt}{0pt} 11/2 & \rule{20pt}{0pt} 15/2 & \rule{20pt}{0pt} 15/2 & \rule{20pt}{0pt} 19/2 & \rule{20pt}{0pt} 19/2 & \rule{20pt}{0pt} 23/2 & \rule{20pt}{0pt} 23/2 \\
        &&&\multicolumn{5}{c}{\rule{20pt}{0pt} $\uparrow\downarrow$ at $p$-wave unitary limit}&&\\ & \rule{20pt}{0pt} \underline{1} & \rule{20pt}{0pt} 7/2 & \rule{20pt}{0pt} 11/2 & \rule{20pt}{0pt} 15/2 & \rule{20pt}{0pt} 15/2 & \rule{20pt}{0pt} 19/2 & \rule{20pt}{0pt} 19/2 & \rule{20pt}{0pt} 23/2 & \rule{20pt}{0pt} 23/2 \\
        &&&\multicolumn{5}{c}{$\uparrow\uparrow$ at $p$-wave unitary limit}&&\\
        & \rule{20pt}{0pt} \underline{1} & \rule{20pt}{0pt} \underline{1} & \rule{20pt}{0pt} 7/2 & \rule{20pt}{0pt} 11/2 & \rule{20pt}{0pt} 15/2 & \rule{20pt}{0pt} 15/2 & \rule{20pt}{0pt} 19/2 & \rule{20pt}{0pt} 19/2 & \rule{20pt}{0pt} 23/2 \\
        &&&\multicolumn{5}{c}{\rule{20pt}{0pt} $\uparrow\downarrow\uparrow$ all at $p$-wave unitary limit}&&\\
        \hline
        \multirow{8}{*}{$1^-$}  & \rule{20pt}{0pt} 5/2 & \rule{20pt}{0pt} 9/2 & \rule{20pt}{0pt} 9/2 & \rule{20pt}{0pt} 13/2 & \rule{20pt}{0pt} 13/2 & \rule{20pt}{0pt} 13/2 & \rule{20pt}{0pt} 17/2 & \rule{20pt}{0pt} 17/2 & \rule{20pt}{0pt} 17/2  \\
        &&&&\multicolumn{3}{c}{\rule{20pt}{0pt} non-interacting}&& \\
        & \rule{20pt}{0pt} \underline{0} & \rule{20pt}{0pt} \underline{2} & \rule{20pt}{0pt} 5/2 & \rule{20pt}{0pt} 9/2 & \rule{20pt}{0pt} 9/2 & \rule{20pt}{0pt} 13/2 & \rule{20pt}{0pt} 13/2 & \rule{20pt}{0pt} 13/2 & \rule{20pt}{0pt} 17/2  \\
        &&&\multicolumn{5}{c}{\rule{20pt}{0pt} $\uparrow\downarrow$ at $p$-wave unitary limit}&&\\
        & \rule{20pt}{0pt} \underline{0} & \rule{20pt}{0pt} \underline{2} & \rule{20pt}{0pt} 5/2 & \rule{20pt}{0pt} 9/2 & \rule{20pt}{0pt} 9/2 & \rule{20pt}{0pt} 13/2 & \rule{20pt}{0pt} 13/2 & \rule{20pt}{0pt} 13/2 & \rule{20pt}{0pt} 17/2 \\
        &&&\multicolumn{5}{c}{\rule{20pt}{0pt} $\uparrow\uparrow$ at $p$-wave unitary limit}&&\\
        & \rule{20pt}{0pt} \underline{0} & \rule{20pt}{0pt} \underline{0} & \rule{20pt}{0pt} \underline{2} & \rule{20pt}{0pt} \underline{2} & \rule{20pt}{0pt} 5/2 & \rule{20pt}{0pt} 9/2 & \rule{20pt}{0pt} 9/2 & \rule{20pt}{0pt} 13/2 & \rule{20pt}{0pt} 13/2  \\
        &&&\multicolumn{5}{c}{\rule{20pt}{0pt} $\uparrow\downarrow\uparrow$ all at $p$-wave unitary limit}&& \\
        \hline
        \multirow{4}{*}{$2^-$}  & \rule{20pt}{0pt} 9/2 & \rule{20pt}{0pt} 9/2 & \rule{20pt}{0pt} 13/2 & \rule{20pt}{0pt} 13/2 & \rule{20pt}{0pt} 13/2 & \rule{20pt}{0pt} 17/2 & \rule{20pt}{0pt} 17/2 & \rule{20pt}{0pt} 17/2 & \rule{20pt}{0pt} 21/2  \\
        &&&&\multicolumn{3}{c}{\rule{20pt}{0pt} non-interacting}&&& \\
        & \rule{20pt}{0pt} \underline{2} & \rule{20pt}{0pt} 9/2 & \rule{20pt}{0pt} 9/2 & \rule{20pt}{0pt} 13/2 & \rule{20pt}{0pt} 13/2 & \rule{20pt}{0pt} 13/2 & \rule{20pt}{0pt} 17/2 & \rule{20pt}{0pt} 17/2 & \rule{20pt}{0pt} 17/2 \\
        &&&\multicolumn{5}{c}{\rule{20pt}{0pt} $\uparrow\downarrow$ at $p$-wave unitary limit}&&\\
        & \rule{20pt}{0pt} \underline{2} & \rule{20pt}{0pt} 9/2 & \rule{20pt}{0pt} 9/2 & \rule{20pt}{0pt} 13/2 & \rule{20pt}{0pt} 13/2 & \rule{20pt}{0pt} 13/2 & \rule{20pt}{0pt} 17/2 & \rule{20pt}{0pt} 17/2 & \rule{20pt}{0pt} 17/2  \\
        &&&\multicolumn{5}{c}{\rule{20pt}{0pt} $\uparrow\uparrow$ at $p$-wave unitary limit}&&\\
        & \rule{20pt}{0pt} \underline{2} & \rule{20pt}{0pt} \underline{2} & \rule{20pt}{0pt} 9/2 & \rule{20pt}{0pt} 9/2 & \rule{20pt}{0pt} 13/2 & \rule{20pt}{0pt} 13/2 & \rule{20pt}{0pt} 13/2 & \rule{20pt}{0pt} 17/2 & \rule{20pt}{0pt} 17/2  \\
        &&&\multicolumn{5}{c}{$\uparrow\downarrow\uparrow$ all at $p$-wave unitary limit}&& \\
        \hline\hline
    \end{tabular}
    \caption{Comparison of the $1^{\text{st}}$ to $9^{\text{th}}$ $l_{e,\nu}$ values characterizing the long range trimer adiabatic potentials for two spin-up and one spin-down fermion $(\downarrow\uparrow\uparrow)$ in different cases.  The first row for each symmetry displayed shows the lowest $l_{e,\nu}$ values that occur for non-interacting particles of each symmetry.  The next three rows for each symmetry give this crucial asymptotic coefficient for different scenarios where some or all $p$-wave interactions are set at $p$-wave unitarity. The values that are underlined integers represent new $l_{e,\nu}$ values that emerge only at the $p$-wave unitary limit.}
    \label{tab:psp}
\end{table}

\twocolumngrid

\section{Three spin-up fermions at $P$-wave unitarity}\label{sec:three p-wave}

\begin{figure}[htbp]
    \centering
    \subfloat[\label{fig:pppVpJ1p}]
    {\includegraphics[width=9cm]{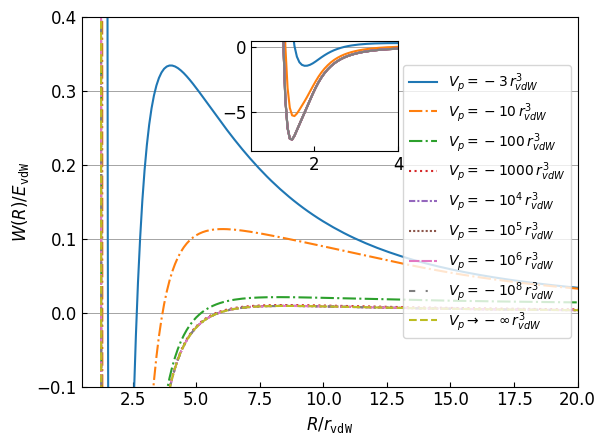}}
    \subfloat[\label{fig:pppVpJ1m}]
    {\includegraphics[width=9cm]{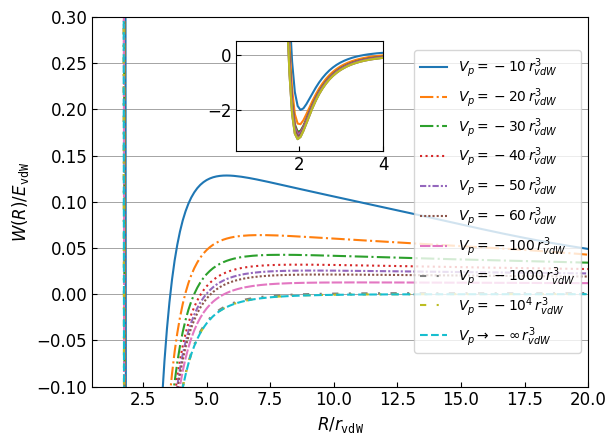}}
    \caption{(color online). Shown are the lowest effective adiabatic potential curves as a function of hyperradius $R$ for several different $p$-wave scattering volumes ($V_p$) in the van der Waals volume unit $r_{\text{vdW}}^3$, for the system $\uparrow\uparrow\uparrow$ in different symmetries $L^\Pi$. As the magnitude of $\lvert V_p \rvert$ gets larger, the potential curve should asymptotically approach the three-body threshold with the parameter characterizing the centrifugal barrier approximately equal to the value $l_e=0$. Inset shows the detail of depth behavior of effective adiabatic potential curves.
    (a), trimer symmetry $L^\Pi=1^+$. (b), trimer symmetry $L^\Pi=1^-$.} 
    \label{fig:pppVpJ1pm}
\end{figure}

Fig.\ref{fig:pppVpJ1pm} compares the lowest effective adiabatic potential curves for three spin-up fermions with different $p$-wave scattering volumes, for both parities and one unit of trimer orbital angular momentum. These two figures demonstrate our finding, namely that the universal trimer energy of three spin-polarized fermions with symmetry $1^+$ is deeper than the case with symmetry $1^-$ because the depth of the lowest effective adiabatic potential curve for $L^\Pi=1^+$ is much deeper than the case with $L^\Pi=1^-$. 

\begin{figure}[htbp]
    \centering
    \subfloat[\label{fig:ppp3rdpwave}]
    {\includegraphics[width=8.5cm]{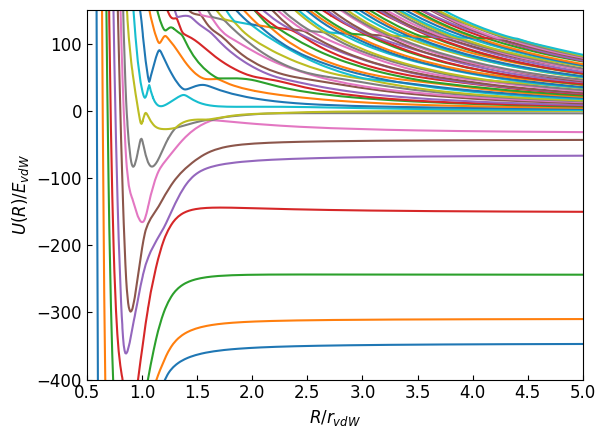}}
    \subfloat[\label{fig:ppp3rdpwaveEig}]
    {\includegraphics[width=9cm]{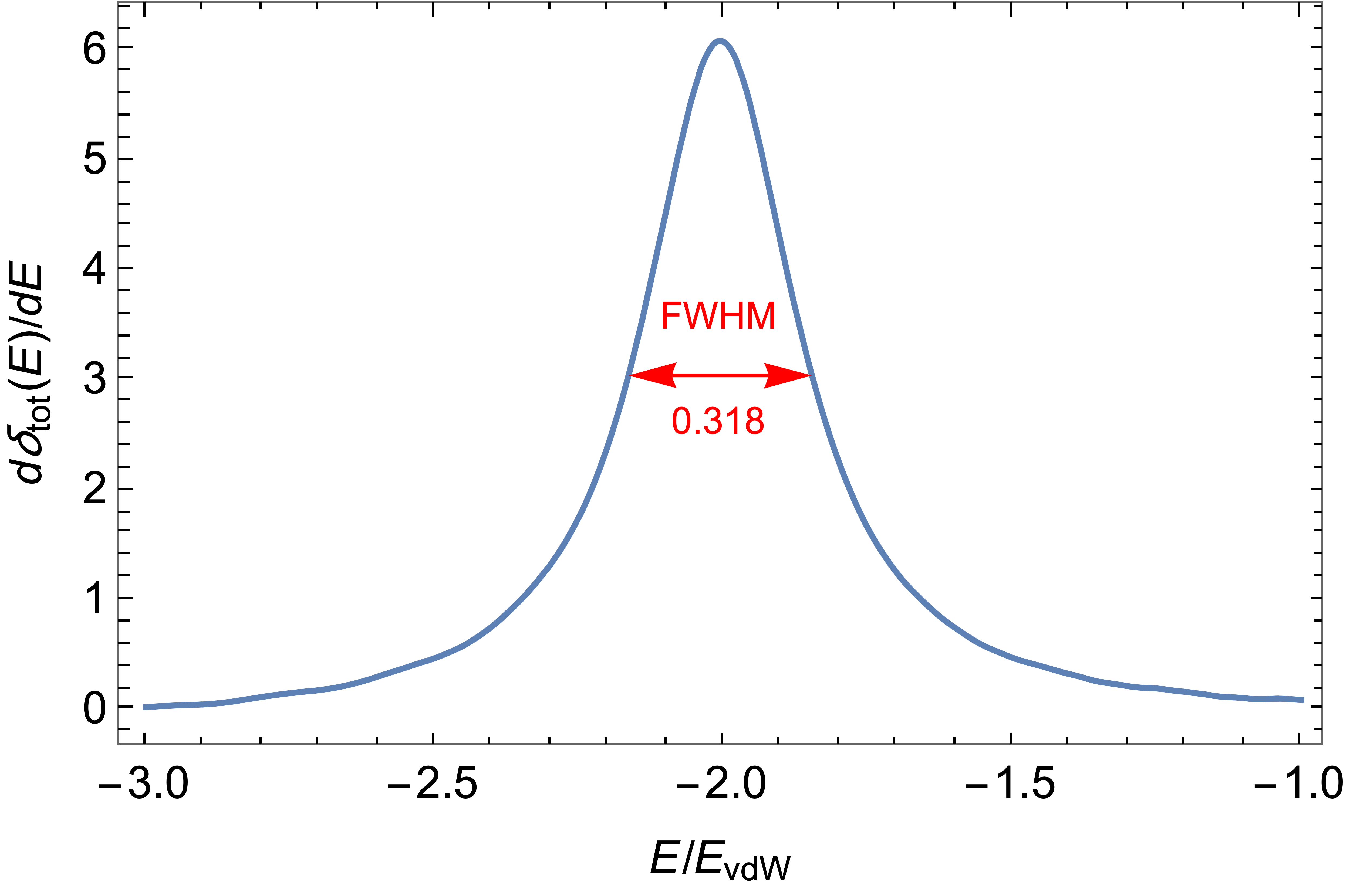}}
    \caption{(color online). (a), Shown are the 1$^{\rm st}$ through 60$^{\rm th}$ three-body Born-Oppenheimer potential curves for three spin-polarized fermions ($\uparrow\uparrow\uparrow$) with in the symmetry $L^{\Pi}=1^+$. The interaction between each pair of fermions has been set at the $3^{\text{rd}}$ $p$-wave pole unitarity which has deep $p$-, $f$- ,$h$-, $j$ and $l$-wave bound states. (b), shown the derivative of the sum of eigenphase shifts as a function of energy which peaks at the position of the universal $p$-wave trimer, around $E=-2.006\, E_{\text{vdW}}$.  The figure indicates the width calculated as the full width at half maximum, namely $\Gamma=0.318 E_{\rm vdW}$.}
    \label{fig:ppp3rdpwave2}
\end{figure}

Table \ref{tab:3Fpppuniversal} shows the $p$-wave trimer universality status for different symmetries, both with and without inclusion of the diagonal $Q_{\nu\nu}(R)$ correction. These results are similar to the previous table for the case of two spin-up and one spin-down fermion, all interactions set at the $p$-wave unitary limit. The interesting ``faux-Efimov'' case occurs for the symmetry $L^\Pi=1^-$ and the universal trimer state exists for symmetries $L^\Pi=1^+$ and $1^-$. Evidence for the universality of the universal trimer is shown in Fig.\ref{fig:ppp3rdpwave2}, which plots the adiabatic potentials for 3 spin-up fermion at the $3^{\text{rd}}$ $p$-wave pole and the derivative of the total eigenphase shift as a function of energy. Fig.\ref{fig:ppp3rdpwaveEig} shows that the universal $p$-wave trimer still exists when the interaction between each pair of fermion is set at the third $p$-wave pole. The corresponding trimer energy is around $E=-2.006\, E_{\text{vdW}}$ which is close to previous results shown as open-circles in Fig.\ref{fig:ppptrimer}.

\begin{figure}[htbp]
    \subfloat[\label{fig:ppp2ndJ1m}]
    {\includegraphics[width=8.5cm]{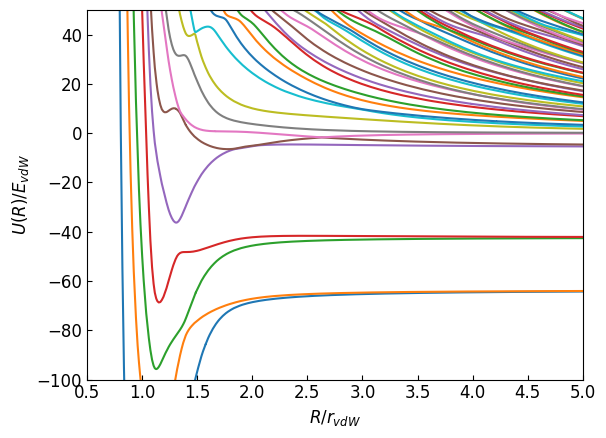}}\centering
    \subfloat[\label{fig:ppp2nddJ1mEig}]
    {\includegraphics[width=10cm]{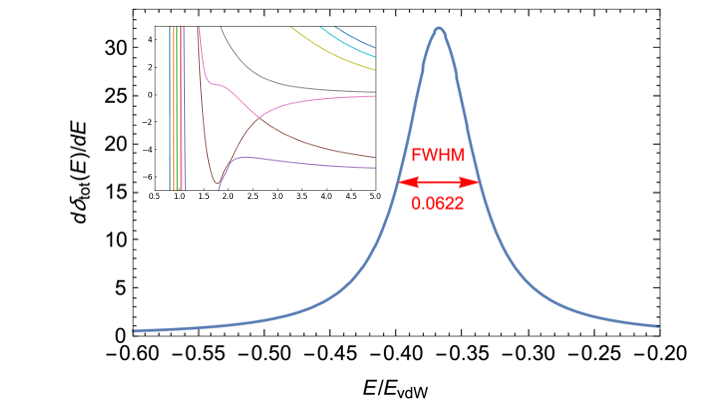}}\\
    \subfloat[\label{fig:ppp3rdJ1m}]
    {\includegraphics[width=8.5cm]{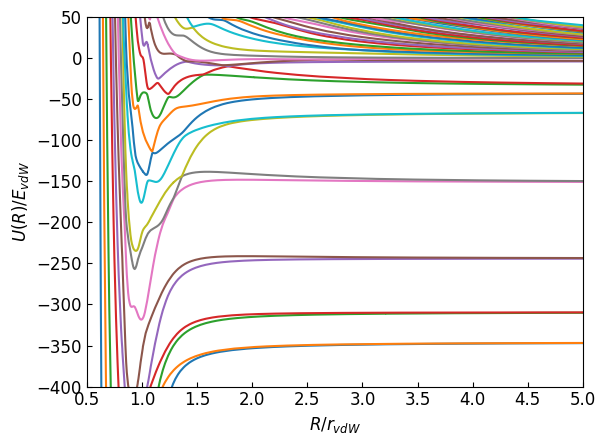}}\centering
    \subfloat[\label{fig:ppp3rdJ1mEig}]
    {\includegraphics[width=10cm]{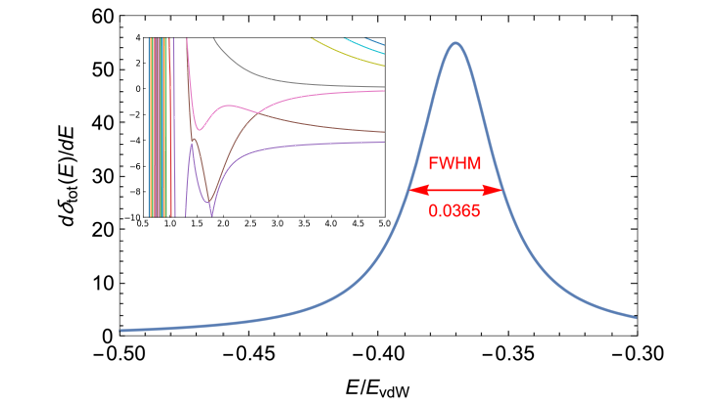}}\\
    \caption{(color online). (a) and (c) Shown are the three-body Born-Oppenheimer potential curves 1-60 for three spin-polarized fermions ($\uparrow\uparrow\uparrow$) with symmetry $L^{\Pi}=1^-$. The interaction between each pair of fermions is set at the $2^{\text{nd}}$ and $3^{\text{rd}}$ $p$-wave pole, respectively. (b) and (d) show the derivative of the sum of eigenphase shifts as a function of energy, and its peak corresponds to the universal $p$-wave trimer resonance energy. The peaks are located around $E=-0.368\, E_{\text{vdW}}$ and $E=-0.370\, E_{\text{vdW}}$ for three spin-up fermion at the $2^{\text{nd}}$ and $3^{\text{rd}}$ $p$-wave pole, respectively. The corresponding resonance widths are indicated on the figure, in vdW energy units.The inset plots in (b) and (d) are zoomed-in views of (a) and (c), respectively, showing the detailed behavior of diabatic potential curves near the $p$-wave unitarity.} 
    \label{fig:ppp2nd3rd}
\end{figure}

In Fig.\ref{fig:pppJ1mtrimer}, the universal trimer energy for two spin-up and one spin-down fermion is higher than the three spin-up fermions (open circles), with values of approximately $E \approx -0.41127\, E_{\text{vdW}}$ and $E \approx -0.40982\, E_{\text{vdW}}$, respectively.  Not surprisingly, there are different non-interacting coefficients of three-body continuum channels, a consequence of the fermionic antisymmetry. (See Table\ref{tab:psp} and Table\ref{tab:ppp} for the non-interacting $l_{e,\nu}$ values.) Performing a 30 channel calculation would produce slightly different universal trimer energies. 
The intersection of the curves with the $x$-axis identifies the point of zero trimer energy, which is the free three-body threshold, and it suggests how an ultracold atomic physics experiment can find the recombination resonance associated with the universal trimer state. Fig.\ref{fig:ppp2nd3rd} shows the evidence that the $p$-wave universal trimer can be found at different $p$-wave poles for three spin-polarized Fermi gases with trimer symmetry $L^\Pi=1^-$. Fig.\ref{fig:ppp2ndJ1m} and \ref{fig:ppp3rdJ1m} plot the lowest 60 Born-Opennheimer potential curves at the $2^{\text{nd}}$ and $3^{\text{rd}}$ $p$-wave pole, respectively. Fig.\ref{fig:ppp2nddJ1mEig} and \ref{fig:ppp3rdJ1mEig} illustrate use of the sum of all eigenphases to obtain the universal trimer (resonance) energies at the $2^{\text{nd}}$ and $3^{\text{rd}}$ $p$-wave pole, respectively. The two energies are very close, which is evidence for the universality of the $p$-wave trimer. Table \ref{tab:ppp} represents the comparison of the first through ninth $l_{e,\nu}$ values for three spin-up fermions at the $p$-wave unitary limit in different symmetries $L^\Pi$. Interestingly, there is a large Efimov reduction of the lowest $l_{e,\nu}$ when $p$-wave two-body interaction goes to a divergent scattering volume for the trimer symmetry $L^\Pi=0^+$, all the way from $l_e=7.5$ (non-interacting) to $l_e=1$ (unitarity). 

\onecolumngrid

\begin{table}[htbp]
    \centering
    \begin{tabular}{ccccc}
    \hline\hline
    \rule{0pt}{10pt} $L^{\Pi}$ & \rule{50pt}{0pt} Trimer Exists? (w/o $Q_{\nu\nu}(R)$) & \rule{50pt}{0pt} Trimer Exists? (w/ $Q_{\nu\nu}(R)$) & \rule{50pt}{0pt} Universality of Trimer? \\
    \hline
    $0^{+}$ & \rule{50pt}{0pt} no  & \rule{50pt}{0pt} no & \rule{50pt}{0pt} no & \\
    $1^{+}$ & \rule{50pt}{0pt} yes & \rule{50pt}{0pt} yes & \rule{50pt}{0pt} yes & \\
    $1^{-}$ & \rule{50pt}{0pt} faux-Efimov & \rule{50pt}{0pt} yes & \rule{50pt}{0pt} yes & \\
    $2^{-}$ & \rule{50pt}{0pt} no & \rule{50pt}{0pt} no & \rule{50pt}{0pt} no & \\
    \hline\hline
    \end{tabular}
    \caption{Comparison of $p$-wave trimer universality status at different symmetries $L^\Pi$ for three spin-up fermions $\uparrow\uparrow\uparrow$ at the $p$-wave unitary limit. The ``faux-Efimov'' case occurs when the lowest Born-Oppenheimer potential curve in the 3-body continuum has a coefficient of $1/R^2$  that is more negative asymptotically than $-1/(8R^2)$ in van der Waals units.
    The ``Universality of Trimer'' column reflects a test of the robustness of the trimer energy obtained at different $p$-wave poles in the two-body interaction potentials.}
    \label{tab:3Fpppuniversal}
\end{table}

\twocolumngrid

\onecolumngrid

\twocolumngrid

\onecolumngrid

\twocolumngrid

\onecolumngrid

\twocolumngrid

\section{Conclusion}

In summary, we have summarized the universal trimer states of three fermion systems with either unequal or equal spin states, in a number of different symmetries. The universal trimer energies have been found for cases where each pair of unequal spin fermion interacts at various $s$-wave and/or $p$-wave poles and each pair of equal spin fermions interact at different $p$-wave poles. The Table \ref{tab:sps}, \ref{tab:psp} and \ref{tab:ppp} illustrate how the $l_{e,\nu}$ values are changed while each pair of fermionic atoms interact at the $s$-wave and/or $p$-wave unitary limit. Interestingly, these results show that the largest Efimov reduction of three fermionic atoms occurs for the case of 3 spin-up fermion at $p$-wave unitarity with $L^\Pi=0^+$, with a difference equal to 6.5. Table \ref{tab:total} summarizes the most significant information from Table \ref{tab:spsuniversal} to Table \ref{tab:ppp}. The $l_{e,\nu}$ value and universal trimer state of all cases of three fermion with equal spin or unequal spin for trimer angular $L^\Pi=0^+$, $1^+$, $1^-$ and $2^-$ is summarized in this table.

\onecolumngrid

\begin{table}[htbp]
    \centering
    \begin{tabular}{c|cccccccccc}
    \hline\hline 
        \rule{0pt}{10pt} $L^\Pi$ &&&&& \rule{20pt}{0pt} $l_{e,\nu}$  &&&&\\
        \hline
        \multirow{4}{*}{$0^+$}  & \rule{20pt}{0pt} 15/2 & \rule{20pt}{0pt} 23/2 & \rule{20pt}{0pt} 27/2 & \rule{20pt}{0pt} 31/2 & \rule{20pt}{0pt} 35/2 & \rule{20pt}{0pt} 39/2 & \rule{20pt}{0pt} 39/2 & \rule{20pt}{0pt} 43/2 & \rule{20pt}{0pt} 47/2   \\
        &&&&\multicolumn{3}{c}{ \rule{20pt}{0pt} non-interacting}&&& \\
        & \rule{20pt}{0pt} \underline{1} & \rule{20pt}{0pt} 15/2 & \rule{20pt}{0pt} 23/2 & \rule{20pt}{0pt} 27/2 & \rule{20pt}{0pt} 31/2 & \rule{20pt}{0pt} 35/2 & \rule{20pt}{0pt} 39/2 & \rule{20pt}{0pt} 39/2 & \rule{20pt}{0pt} 43/2 \\
        &&&\multicolumn{5}{c}{\rule{20pt}{0pt} $\uparrow\uparrow\uparrow$ at $p$-wave unitary limit}&& \\ 
        \hline
        \multirow{4}{*}{$1^+$}  & \rule{20pt}{0pt} 7/2 & \rule{20pt}{0pt} 15/2 & \rule{20pt}{0pt} 19/2 & \rule{20pt}{0pt} 23/2 & \rule{20pt}{0pt} 27/2 & \rule{20pt}{0pt} 31/2 & \rule{20pt}{0pt} 31/2 & \rule{20pt}{0pt} 35/2 & \rule{20pt}{0pt} 39/2   \\
        &&&&\multicolumn{3}{c}{\rule{20pt}{0pt} non-interacting}&& \\
        & \rule{20pt}{0pt} \underline{1} & \rule{20pt}{0pt} 7/2 & \rule{20pt}{0pt} 15/2 & \rule{20pt}{0pt} 19/2 & \rule{20pt}{0pt} 23/2 & \rule{20pt}{0pt} 27/2 & \rule{20pt}{0pt} 31/2 & \rule{20pt}{0pt} 31/2 & \rule{20pt}{0pt} 35/2 \\
        &&&\multicolumn{5}{c}{\rule{20pt}{0pt} $\uparrow\uparrow\uparrow$ at $p$-wave unitary limit}&& \\
        \hline
        \multirow{4}{*}{$1^-$}  & \rule{20pt}{0pt} 9/2 & \rule{20pt}{0pt} 13/2 & \rule{20pt}{0pt} 17/2 & \rule{20pt}{0pt} 21/2 & \rule{20pt}{0pt} 21/1 & \rule{20pt}{0pt} 25/2 & \rule{20pt}{0pt} 25/2 & \rule{20pt}{0pt} 29/2 & \rule{20pt}{0pt} 29/2  \\
        &&&&\multicolumn{3}{c}{\rule{20pt}{0pt} non-interacting}&& \\
        & \rule{20pt}{0pt} \underline{0} & \rule{20pt}{0pt} \underline{2} & \rule{20pt}{0pt} 9/2 & \rule{20pt}{0pt} 13/2 & \rule{20pt}{0pt} 17/2 & \rule{20pt}{0pt} 21/2 & \rule{20pt}{0pt} 21/2 & \rule{20pt}{0pt} 25/2 & \rule{20pt}{0pt} 25/2 \\
        &&&\multicolumn{5}{c}{\rule{20pt}{0pt} $\uparrow\uparrow\uparrow$ at $p$-wave unitary limit}&& \\
        \hline
        \multirow{4}{*}{$2^-$}  & \rule{20pt}{0pt} 13/2 & \rule{20pt}{0pt} 17/2 & \rule{20pt}{0pt} 21/2 & \rule{20pt}{0pt} 25/2 & \rule{20pt}{0pt} 25/2 & \rule{20pt}{0pt} 29/2 & \rule{20pt}{0pt} 29/2 & \rule{20pt}{0pt} 33/2 & \rule{20pt}{0pt} 33/2  \\
        &&&&\multicolumn{3}{c}{\rule{20pt}{0pt} non-interacting}&&& \\
        & \rule{20pt}{0pt} \underline{2} & \rule{20pt}{0pt} 13/2 & \rule{20pt}{0pt} 17/2 & \rule{20pt}{0pt} 21/2 & \rule{20pt}{0pt} 25/2 & \rule{20pt}{0pt} 25/2 & \rule{20pt}{0pt} 29/2 & \rule{20pt}{0pt} 29/2 & \rule{20pt}{0pt} 33/2 \\
        &&&\multicolumn{5}{c}{\rule{20pt}{0pt} $\uparrow\uparrow\uparrow$ at $p$-wave unitary limit}&& \\
        \hline\hline
    \end{tabular}
    \caption{Comparison of the $1^{\text{st}}$ through $9^{\text{th}}$ $l_{e,\nu}$ values for three spin-up fermions $(\uparrow\uparrow\uparrow)$ for both the non-interacting limit and for the case where each pair of fermions interacts at $p$-wave unitarity. The red (integer) number represents the new $l_{e,\nu}$ value that emerges at the $p$-wave unitary limit.}
    \label{tab:ppp}
\end{table}

\begin{table}
    \begin{tabular}{cccccc}
    \hline\hline
    \rule{0pt}{10pt}$L^{\Pi}$ & \rule{35pt}{0pt} $l_{e,\nu}$ & \rule{35pt}{0pt} Trimer (w/o $Q_{\nu\nu}$) & \rule{35pt}{0pt} Trimer (w/ $Q_{\nu\nu}$) & \rule{35pt}{0pt} Universality of Trimer \\
    &&&& \\
    \hline
    $0^{+}$ &\rule{20pt}{0pt} $\underline{1}$, $\textbf{1.666}$, $\textbf{4.628}$, $\textbf{6.615}$, $7/2$ & \rule{20pt}{0pt} no & \rule{20pt}{0pt} no & \rule{20pt}{0pt} no &\\
    $1^{+}$ &\rule{20pt}{0pt} $\underline{1}$, $7/2$, $11/2$, $15/2$, $15/2$ &\rule{20pt}{0pt} yes & \rule{20pt}{0pt} no & \rule{20pt}{0pt} no &\\
    $1^{-}$ &\rule{20pt}{0pt} $\underline{0}$, $\textbf{1.272}$, $2$, $\textbf{3.858}$, $9/2$ & \rule{20pt}{0pt}faux-Efimov & \rule{20pt}{0pt} yes & \rule{20pt}{0pt} yes &\\
    $2^{-}$ & \rule{20pt}{0pt} $\underline{2}$, $9/2$, $9/2$, $13/2$, $13/2$ &\rule{20pt}{0pt}no & \rule{20pt}{0pt} no & \rule{20pt}{0pt} no &\\
    &\multicolumn{4}{c}{\rule{20pt}{0pt} (a) $\uparrow\downarrow$ at $s$-wave unitarity and $\uparrow\uparrow$ at $p$-wave unitarity limit}\\
    $0^{+}$ &\rule{20pt}{0pt} $\underline{1}$, $7/2$, $11/2$, $15/2$, $15/2$ &\rule{20pt}{0pt}  no &\rule{20pt}{0pt} no &\rule{20pt}{0pt} no &\\
    $1^{+}$ &\rule{20pt}{0pt} $\underline{1}$, $7/2$, $11/2$, $15/2$, $15/2$ &\rule{20pt}{0pt} yes & \rule{20pt}{0pt} yes & \rule{20pt}{0pt} yes &\\
    $1^{-}$ &\rule{20pt}{0pt} $\underline{0}$, $\underline{2}$, $5/2$, $9/2$, $9/2$ &\rule{20pt}{0pt} faux-Efimov &\rule{20pt}{0pt} no & \rule{20pt}{0pt} no &\\
    $2^{-}$ &\rule{20pt}{0pt} $\underline{2}$, $9/2$, $9/2$, $13/2$, $13/2$ & \rule{20pt}{0pt} no & \rule{20pt}{0pt} no & \rule{20pt}{0pt} no &\\
    &\multicolumn{4}{c}{\rule{20pt}{0pt} (b) $\uparrow\downarrow$ at $p$-wave unitarity and
    $\uparrow\uparrow$ with weak interaction}\\
    $0^{+}$ &\rule{20pt}{0pt}  $\underline{1}$, $\underline{1}$, $7/2$, $11/2$, $15/2$ & \rule{20pt}{0pt} no & \rule{20pt}{0pt} no & \rule{20pt}{0pt} no &\\
    $1^{+}$ &\rule{20pt}{0pt}  $\underline{1}$, $\underline{1}$, $7/2$, $11/2$, $15/2$ & \rule{20pt}{0pt} yes & \rule{20pt}{0pt} yes & \rule{20pt}{0pt} yes &\\
    $1^{-}$ &\rule{20pt}{0pt} $\underline{0}$, $\underline{0}$, $\underline{2}$, $\underline{2}$, $5/2$ & \rule{20pt}{0pt} faux-Efimov*2 & \rule{20pt}{0pt} yes & \rule{20pt}{0pt} yes &\\ 
    $2^{-}$ &\rule{20pt}{0pt} $\underline{2}$, $\underline{2}$, $9/2$, $9/2$, $13/2$ & \rule{20pt}{0pt} no & \rule{20pt}{0pt} no & \rule{20pt}{0pt} no &\\
    &\multicolumn{4}{c}{\rule{20pt}{0pt} (c) $\uparrow\downarrow\uparrow$ all at $p$-wave unitarity}\\
    $0^{+}$ &\rule{20pt}{0pt} $\underline{1}$, $15/2$, $23/2$, $27/2$, $31/2$ & \rule{20pt}{0pt} no & \rule{20pt}{0pt} no & \rule{20pt}{0pt} no &\\
    $1^{+}$ &\rule{20pt}{0pt} $\underline{1}$, $7/2$, $15/2$, $19/2$, $23/2$ & \rule{20pt}{0pt} yes & \rule{20pt}{0pt} yes & \rule{20pt}{0pt} yes &\\
    $1^{-}$ & $\rule{20pt}{0pt} \underline{0}$, $\underline{2}$, $9/2$, $13/2$, $17/2$ & \rule{20pt}{0pt} faux-Efimov & \rule{20pt}{0pt} yes & \rule{20pt}{0pt} yes &\\
    $2^{-}$ & \rule{20pt}{0pt} $\underline{2}$, $13/2$, $17/2$, $21/2$, $25/2$ & \rule{20pt}{0pt} no & \rule{20pt}{0pt} no & \rule{20pt}{0pt} no &\\
    &\multicolumn{4}{c}{\rule{20pt}{0pt} (d) $\uparrow\uparrow\uparrow$ all at $p$-wave unitarity}\\
    \hline\hline
    \end{tabular}
    \caption{ Comparison of the properties  of two-component or one spin component fermionic trimers at $s$- and/or $p$-wave unitarity for various symmetries $L^{\Pi}$. Shown are the $l_{e,\nu}$ values from the lowest adiabatic channel to the fifth adiabatic channel. The integer $l_{e,\nu}$ values (underlined) represent the $p$-wave unitary channels, the non-integer and non-half-integer $l_{e,\nu}$ values (bold) represent the $s$-wave unitary channels, and the half-integer $l_{e,\nu}$ values are the same as the values for noninteracting(NI) three particles asymptotically ($R\rightarrow\infty$). The third to fifth column show the assessment of the universality of trimer states with and without the non-adiabatic diagonal $Q_{\nu\nu}(R)$ correction. (a) The interaction between different spin fermions is set close to the $s$-wave unitary limit, and the interaction between same spin fermions is set close to the $p$-wave unitary limit. (b) The interaction between spin-up and spin-down fermions is set at the $p$-wave unitary limit and with weak interaction between the pair of spin-up fermions. (c) The interaction between each pair of spin-polarized fermions or opposite spin fermions are all set at $p$-wave unitarity. (d) The interaction between each pair of spin-polarized fermions are all set at $p$-wave unitarity.}
    \label{tab:total}
\end{table}

\twocolumngrid

\section*{Acknowledgement}

This work was supported in part by NSF Grant Award Number 2207977 and in part by the National Science and Technology Council of Taiwan, with grant numbers NSTC 113-2112-M-019-003-MY3. We thank Jia Wang and Jose D'Incao for sharing their computer programs and Michael Higgins for discussions.

\bibliography{reference}

\end{document}